\documentclass{jfm}
\usepackage{bm} 
\usepackage{float} 

\usepackage{fancyhdr}
\title{Self-focusing of helicity drives finite-time singularities  in inviscid flows}
\author{Mokhtar Adda-Bedia\aff{1} \and Sergio Rica\aff{2}}
\affiliation{\aff{1}Laboratoire de Physique, CNRS, ENS de Lyon, Universit\'e de Lyon, F-69364 Lyon, France
\aff{2}Instituto de F\'isica, Facultad de F\'isica, Pontifica Universidad Cat\'olica de Chile, Casilla 306, Santiago,  Chile}

\begin{document}
\maketitle

\begin{abstract}
This paper deals with the longstanding quest of the possible existence of finite-time singularities in the equations governing the dynamics of inviscid  fluids, namely, Euler equations. Here, two contributions are brought for the case of perfect fluids with finite initial energy. First, a self-similar velocity field inspired by Leray Ansatz is proposed which allows for a separation of variables that transforms the original partial differential Euler equations to a nonlinear system of ordinary differential equations. This system can be solved semi-analytically and allows a continuum set of solutions parametrised by a self-similar exponent, $\nu$. Second, we use the conservation laws of Euler equations to select the possible finite-time singular solutions  and the related self-similar exponents. We find that the helicity is the driving mechanism of the blow-up through a self-focusing mechanism. The flow near the singularity separates into two phases. A first phase is within a tubular region that shrinks as a power-law $(t_c-t)^\nu$, with $t_c$ the blow-up time, where the  helicity is focused. This region is separated by a sharp interface from an outer region where the vorticity, and thus helicity, is identically zero. We found that the finite-time singularity may be either point-like or line-like depending on the dynamics of the tubular region along its axis of symmetry. Incidentally for a point-like singularity we recover the Leray scaling $\nu=1/2$ paving the way to a generalisation of this approach for the Navier-Stokes equations. Finally, we conjecture that if the helicity vanishes initially, no finite-time singularity would be possible, since in this case the singularity occurs at infinite time from the initial condition.
\end{abstract}


\section{Introduction}

More than 30 years ago, \cite{yves1994} has suggested a mechanism for the self-focusing of vorticity as a columnar vortex filament in an inviscid and incompressible fluid in connection with experimental observations of \cite{Douady1991} and the results of numerical simulations of \cite{Brachet1991}. Pomeau regarded this mechanism as a consequence of a potential finite time blow-up of the solution of Euler equations with axial symmetry in three space dimensions. 

Despite a quarter of a millennium of history,  the nature of Euler equations solutions remains an open problem. For instance, to date, it is not known whether, for smooth initial conditions of finite energy, the solutions of Euler equations or their gradients explode in finite time or not. The problem of the existence of singular solutions of incompressible Euler equations dates back at least to the 1920s and 30s~\citep{Lichtenstein,Gunther} and the seminal work of \cite{leray1934} on the nature of solutions of the Navier-Stokes equations\footnote{A problem  which is related to the sixth millennium problem proposed by the Clay Institute: https://www.claymath.org/millennium/navier-stokes-equation/}. The question of regularity regained interest in the 80s from a numerical and theoretical perspective. Both, physicists and mathematicians have made an enormous effort to prove or disprove the regularity problem~\citep{Gibbon08}. 

 More recently, the existence or non existence of singular finite-time solutions of Euler equations has witnessed an important development since the numerical evidence by \cite{Luo2014} of the existence of a self-similar blow-up solution for the axial vorticity arising at the boundary of a confined axial flow \citep{Barkley2020}.  This line-like singularity was  studied and confirmed numerically by a study using PINNs \citep{Buckmaster}, and was also proven mathematically by~\cite{ChenHou2025}. 

On the other hand, point-like singularities were suggested by Pomeau in a series of papers~\citep{yves1994,yves1995,yves2018}, in particular in connection with the anomalous dissipation in fully developed turbulence~\citep{yves2019,martine19,PRFJoss2020,PhysD2023}. An explicit approximation for an axisymmetric flow was proposed by \cite{Elgindi2021annals} who suggested  the existence of a finite-time blow-up of vorticity for a zero swirl flow. This singularity sets up only because of a non-smooth initial condition of the vorticity and the self-induced flow~\citep{Ukhovskii1968}.

Lastly,  following the footprints of \cite{Elgindi2021annals} and by assuming an axi-symmetric flow with swirl that obeys a self-similarity of second type with an unknown exponent, $\nu$, it was shown that Euler equations can be mapped into a dynamical system of an infinite set of ordinary differential equations (ODEs) governing the amplitudes of the relevant fields expanded in spherical harmonics~\citep{CMR2023,PRF2024Diego}. The nonlinear system of ODEs is ultimately solved numerically by truncating the dynamical system up to a certain number of modes leading to a convergent scenario for the solutions and for the nonlinear eigenvalue $\nu$~\citep{PRF2024Diego}. However, this method presents a relevant failure: nonlinear dynamical systems are usually sensitive to initial conditions and this sensitivity is enhanced as one increases the number of modes. The solutions have shown unavoidable oscillations near the fixed points of the dynamical system that characterises the self-similar solutions. Nevertheless, by filtering these oscillations, the ratio of convergence for the value of the self-similar exponent was estimated $\nu\approx 2.03$, presenting a promising path for a numerical or analytical approach indicating a possible $\nu= 2$. The question of a point-like finite-time singularity as a consequence of the evolution by Euler equations in an infinite domain for a smooth and finite energy initial condition remains widely open.

\begin{figure} 
\centerline{\includegraphics[width=12cm]{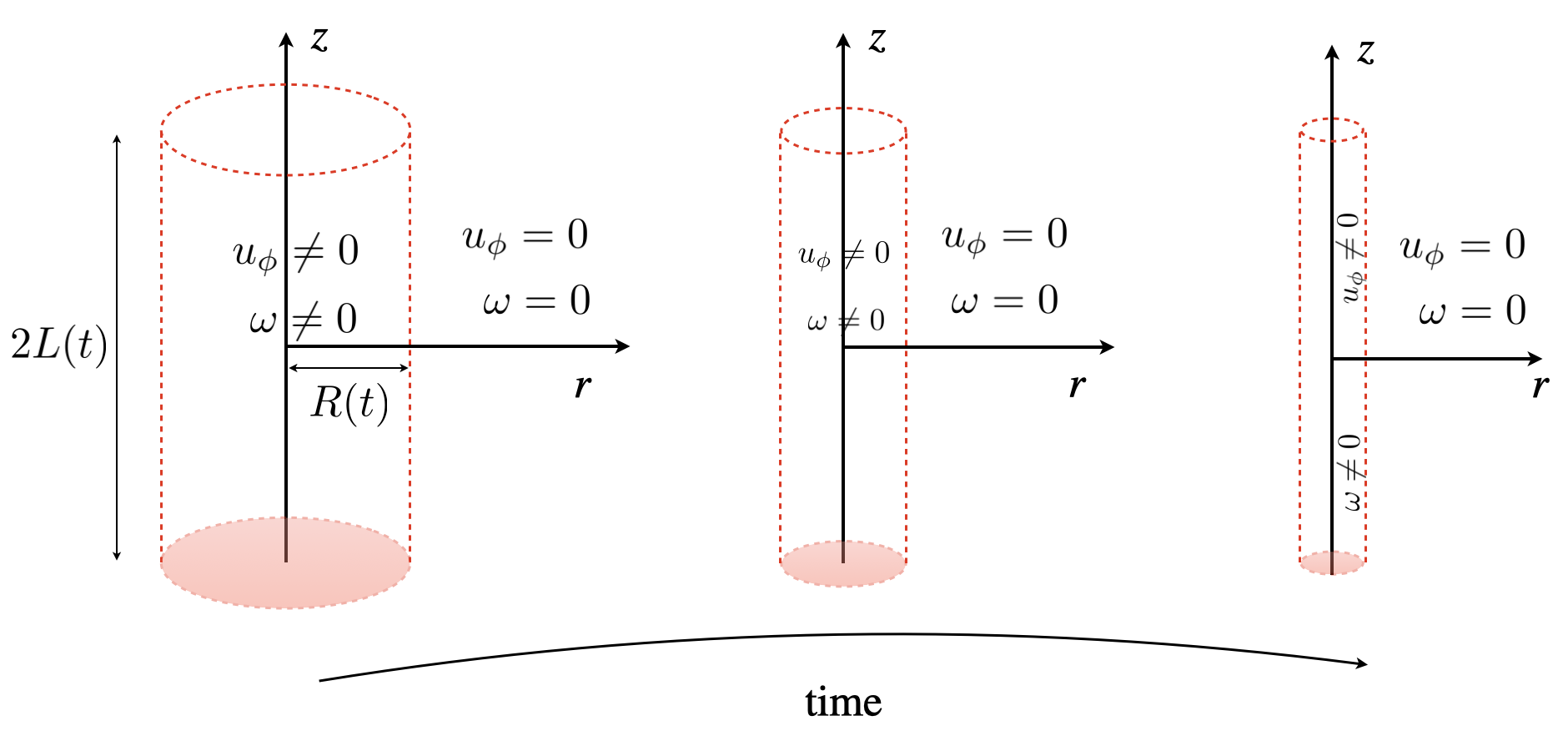}}
\caption{ \label{Fig:Fig1} Scheme for the finite-time singular solution. Inside the cylindrical region delimited by a radius $R(t)$ the swirl velocity, $u_\phi$, and vorticity, $\omega_\phi$, are not zero, while outside this region both vorticity and swirl velocity vanish.  As we will see,  an adequate separation of variables leads to $R(t)  \to 0 $ in finite time. Depending on the dynamics of $L(t)$, the tubular region shrinks in finite time into a point or into a line. }
\end{figure}

In this article, we consider a special kind of self-similar solution which places in advantage the original idea by \cite{yves1994} considering a columnar distribution of vorticity. Briefly, we show the existence of a solution of Euler equations that focuses both vorticity and swirl velocity inside a delimited cylindrical region characterised by $R(t)$. This idea is sketched in Fig.~\ref{Fig:Fig1} based on the results presented {\it cf. infra}. Outside this cylindrical region the vorticity and the swirl velocity vanish. The radius $R(t)$ contracts and eventually vanishes in finite time. Notice that the distribution of vorticity proposed here makes the vorticity flux, and thus, the circulation in the vertical direction exactly zero, $\Gamma_\phi = \int \omega_z  \, r dr d\phi=0$, therefore, it does not correspond to a process of focusing of a net vorticity into a filamentary vortex with finite circulation \citep{Saffman1995vortex}, because, precisely, for a vortex filament  the long range behaviour of the swirl velocity does not vanish. In the current case the helicity is exclusively focused into a tubular column. Nevertheless, we emphasise that because swirl velocity vanishes in a not differentiable manner, the solution proposed here is just $C^0$. 

This paper is organised as follow. In Section~\ref{Secc:Formulation} the formulation of the problem is introduced. Then, in Section~\ref{Secc:Ansatz}, we provide an original Ansatz for the self-similar flow and determine the self-similar Euler-Leray equations. Then, in Sections~\ref{Secc:quatre} and~\ref{Secc:Swirled}, we explore the solutions of the corresponding dynamical system where we uncover  analytical and semi-analytical solutions. Then, in Section~\ref{Secc:Conserved}, we confront these solutions to the conservation laws, which allows us to determine the properties of the flow near the possible singularities including the self-similar exponents. Finally, we end the paper by a discussion of our results and possible extensions.

\section{Formulation of the problem}
\label{Secc:Formulation}

Written more than a quarter of millennium ago~\citep{Euler}, Euler equations are one of the most fundamental equations of fluid mechanics and belong to the oldest partial differential equations. Today they are considered to be relevant in a myriad of situations of everyday life of science and technology~\citep{anderson1995computational}. Nevertheless, they still hide some mysteries. Among them, the question: Does a flow with smooth and finite energy initial conditions in the whole volume, $\mathbb R^3$, remains smooth or blows-up in finite-time~?

Euler equations read
\begin{eqnarray}
\frac{\partial \bm{u}}{\partial t}+\left(\bm{u}\cdot\bm{\nabla}\right)\bm{u}&=&-\frac{1}{\rho}\bm{\nabla} p\;,\label{eq:Euler} \\
\bm{\nabla}\cdot\bm{u}&=&0\;\label{eq:Continuity}
\end{eqnarray}
where Eq.~(\ref{eq:Euler}) corresponds to Newton's second law for a fluid element under the influence of the pressure field, $p( {\bm x},t) $, that takes into account the internal forces acting on that element, and Eq.~(\ref{eq:Continuity}) represents the incompressibility condition. Without loss of generality, we take $\rho=1$ in the subsequent calculations. Contrary to Navier-Stokes equations which provide a description for ordinary viscous fluids, Euler equations are for perfect or inviscid fluids and do not involve any intrinsic physical length scale. Finally, the nonlinear and nonlocal partial differential equations (\ref{eq:Euler}) and (\ref{eq:Continuity}) are complemented by a smooth divergence-free initial condition 
\begin{equation}
{\bm u} ( {\bm x},t=0) ={\bm u}_0 ( {\bm x})\neq0\; .
\end{equation}

Euler equations are formally a reversible and conservative evolution system. In three space dimensions, they possess various conserved quantities, among them the kinetic energy,
\begin{equation}
{\mathcal E}[\bm u] = \frac{1}{2} \int {\bm u}^2 d^3 {\bm x}  > 0\;,
\label{eq:Energy}
\end{equation}
and the helicity, 
\begin{equation}
{\mathcal H} [\bm u]  =\int {\bm u}\cdot ({\bm \nabla} \times {\bm u})\, d^3 {\bm x} \;.
\label{eq:Helicity}
\end{equation}
In what follows, we limit only to finite energy and helicity solutions, that is ${\mathcal E}[{\bm u} ]  < \infty$ and $\left| {\mathcal H}[{\bm u} ] \right| < \infty$. Therefore  $\forall t, \, | {\bm u} ( {\bm x},t)|\to 0 $ fast enough as $| {\bm x}|\to \infty$. We underline that infinite energy blow-up solutions exist in the literature \citep{Gibbon_2003}. However, the authors are not aware if such result exists for infinite helicity initial conditions.

We end this section by writing Euler equations in cylindrical coordinates assuming an axisymmetric flow, that is, the fields do not depend on the angular variable $\phi$. The Euler  equations are simplified into
\begin{eqnarray}
\frac{\partial u_r}{\partial t}+ u_r \frac{\partial u_r}{\partial r}
+ u_z \frac{\partial u_r}{\partial z}- \frac{u_\phi^2}{r}
&=& - \frac{\partial p}{\partial r}\;, \label{eq:ur}\\
\frac{\partial u_z}{\partial t} + u_r \frac{\partial u_z}{\partial r} + u_z \frac{\partial u_z}{\partial z} &=& -\frac{\partial p}{\partial z}\;, \label{eq:uz}\\
\frac{\partial u_\phi}{\partial t} + u_r \frac{\partial u_\phi}{\partial r} + u_z \frac{\partial u_\phi}{\partial z} + \frac{u_r u_\phi}{r} &= &0\;.\label{eq:uphi}
\end{eqnarray}
Before going further, notice that the azimuthal momentum balance equation (\ref{eq:uphi}) is linear in swirl velocity $u_\phi$. 

Finally, the incompressibility condition (\ref{eq:Continuity}) reads
\begin{equation}
\frac{1}{r}\frac{\partial (r u_r)}{\partial r}+ \frac{\partial u_z}{\partial z} = 0\;.\label{eq:Divu}
\end{equation}
In what follows we restrict to solutions having an odd symmetry regarding the $z$-variable, in such a way that $u_z$ and $u_\phi$ are odd functions in $z$ and $u_r$ is an even function of $z$. On the other hand, Equations~(\ref{eq:ur},\ref{eq:uz},\ref{eq:uphi},\ref{eq:Divu})  must be well defined at the axis $r=0$. Therefore, these equations are restricted to the domain $0\leq r<\infty$ and $0\leq z<\infty$ with the boundary conditions
\begin{eqnarray}
u_r \left(r=0,z\right) =u_\phi\left(r=0,z\right) =\frac{\partial u_z}{ \partial r} \left(r=0,z\right) &=& 0\;,\label{eq:BC1} \\
u_z \left(r,z=0\right)=u_\phi \left(r,z=0\right)&=& 0 \; .\label{eq:BC2} 
\end{eqnarray}
The vorticity components for an axisymmetric flow are given by
\begin{equation}
\omega_r =  - \frac{\partial u_\phi}{\partial z} ,\quad \omega_\phi = \frac{\partial u_r}{\partial z} - \frac{\partial u_z}{\partial r}, \quad \omega_z = \frac{1 }{r}   \frac{\partial \left( ru_\phi\right)}{\partial r} \;.\label{eq:Vorticity}
\end{equation}
The regularity of the vorticity components along the $z$-axis imposes $\omega_\phi=0$, which is consistent with the boundary condition $\frac{\partial u_z}{ \partial r} \left(r=0,z\right)=0$ in (\ref{eq:BC1}). Accordingly, $\omega_r$ also vanishes at the axis. On the contrary, the $\omega_z$ component is not constrained along the $z$-axis.

\section{Leray solutions and self-similar equations}
\label{Secc:Ansatz}

\subsection{Self-similar Ansatz}
\label{Secc:Ansatz1}

Instead of searching for an arbitrary singularity, we focus on a special kind of self-similar singularity that has been the center of research over the last 25 years \citep{Barenblatt_1996,eggers_fontelos_2015}. Following the footprints of \cite{leray1934} and \cite{yves1995},  we suggest a  line-like singularity solution of the  fluid dynamics equations in the form
\begin{eqnarray}
u_r(r,z,t) &=& \frac{r}{t_c-t }U_r \left( \frac{r}{R(t)  } \right)\;, \label{eq:AnsatzLerayUr}\\
u_z(r,z,t) &=& \frac{z}{t_c-t }U_z \left( \frac{r}{R(t)  } \right)\;, \label{eq::AnsatzLerayUz}\\
u_\phi(r,z,t)& = &\frac{z}{t_c-t }U_\phi \left( \frac{r}{R(t)  } \right)\;, \label{eq::AnsatzLerayUphi}\\
 p (r,z,t) &=& \frac{r^2 }{(t_c-t)^{2} } {\mathcal P}_r \left( \frac{r}{R(t)  } \right) + \frac{z^2 }{(t_c-t)^{2} } {\mathcal P}_z \left( \frac{r}{R(t)  } \right) \;,
\label{eq::AnsatzLerayPressure}
\end{eqnarray}
where $ U_r(\xi )$, $U_z(\xi )$, $U_\phi(\xi), {\mathcal P}_r(\xi )$  and ${\mathcal P}_z(  \xi )$ are the self-similar velocity  and pressure functions that depend only on the self-similar radial variable: ${ \xi}= r/R(t)$. Here, we have exploited the dimensionalities and parity conditions of the fields. The problem involves a velocity and a pressure per unit of mass density whose dimensions can be built upon the space and time coordinates $\bm{u}\sim \bm{x}/t$ and $p\sim  x^2/t^2$.  When combined with the symmetry conditions of the different fields with respect to the plane $ z = 0 $, the proposed Ansatz is a sufficiently general and minimal form that allows for the separation of the contributions of the spatial variables $r$ and $z$ for each field. Finally, as shown in  Eqn. (\ref{eq:ForR'}) {\it cf. infra}, the function $R(t) $ is required to follow a power law given by
\begin{eqnarray}
R(t) &=&R_0 \left( \frac{t_c-t}{t_c } \right)^\nu\;, \label{eq:SelfSimilarRadiius}
\end{eqnarray}
where $R_0$ is a length scale and $t_c $ is the blow-up time. Both quantities depend on initial conditions and cannot be determined under this simplified Ansatz. However, $R_0$ and $t_c$ should depend on the initial values of the conserved kinetic energy $\mathcal{E}_0$ and helicity $\mathcal{H}_0$. By a dimensional argument and using Equations~(\ref{eq:Energy},\ref{eq:Helicity}), one obtains the following scaling relations
\begin{equation}
\mathcal{E}_0\sim R_0^5 t_c^{-2} \quad \& \quad \mathcal{H}_0\sim R_0^4t_c^{-2}\; . 
\label{eq:scales}
\end{equation}
On the contrary,  $\nu$ is an unknown exponent that should be determined by the explicit solution of the problem. This  problem is known as a second type of self-similarity in the classification of \cite{Zeldovich} and \cite{Barenblatt_1996}. For Navier-Stokes equations, as shown by \cite{leray1934}, a dimensional argument leads to $R(t) =\sqrt{\eta(t_c-t)} $, where $\eta$ is the kinematic viscosity.

Finally, the vorticity components have the right $1/(t_c-t)$ dependence near the singularity, and the right parity regarding $z=0$~:
\begin{equation}
\omega_r =  - \frac{1}{t_c-t }U_\phi \left( \xi \right)\;,\quad \omega_\phi =  -  \frac{1}{t_c-t }  \frac{z}{R(t) }  U_z'(\xi)\;, \quad \omega_z =  \frac{1}{t_c-t }  \frac{z }{R(t)}  \frac{1 }{\xi }   \frac{d \left( \xi U_\phi\right)}{d \xi} \; .\label{eq:VorticitySelf}
\end{equation}
Notice that the dimensionless length $z/R(t) $ in $\omega_\phi$ and $\omega_z$ in Eq. (\ref{eq:VorticitySelf}) diverges $\left(z/R(t) \to \infty\right)$ as $t\to t_c$.

\subsection{Self-similar equations}

Introducing the Ansatz (\ref{eq:AnsatzLerayUr},\ref{eq::AnsatzLerayUz},\ref{eq::AnsatzLerayUphi},\ref{eq::AnsatzLerayPressure})  into the Euler equations (\ref{eq:ur},\ref{eq:uz},\ref{eq:uphi},\ref{eq:Divu}), one first obtains after a direct calculation
\begin{eqnarray}
(t_c-t) \frac{\dot R}{ R} = - \nu\;, \label{eq:ForR'}
\end{eqnarray}
whose solution is given by Eq.~(\ref{eq:SelfSimilarRadiius}). Second, one shows that the variable $ {\mathcal P}_r(\xi )$ satisfies
\begin{equation}
 \xi  {\mathcal P}_r'(\xi )+2  {\mathcal P}_r(\xi ) +( U_r(\xi ) +\nu)  \xi  U_r'(\xi )+U_r(\xi )^2+U_r(\xi ) =0 \;, 
\end{equation}
which, provided regularity conditions at $\xi=0$, can be computed exactly as function of $U_r(\xi)$~:
\begin{equation}
{\mathcal P}_r(\xi ) = -\frac{1}{2} U_r(\xi)^2  -\nu U_r(\xi ) - \frac{(1-2 \nu)}{\xi^2} \int_0^\xi U_r(s) s ds .
\label{eq:LerayPrBis}
\end{equation}
On the other hand, the variables, $U_r$, $U_z$, $U_\phi$ and ${\mathcal P}_{z}$ obey the following ordinary differential equations 
\begin{eqnarray}
  \xi U'_r(\xi)&=&-2 U_r(\xi)-U_z(\xi)\;,\label{eq:LerayDiv}\\
  \xi U'_z(\xi )&=&-\frac{2{\mathcal P}_z(\xi)+U_z(\xi)^2+U_z(\xi)}{\nu +U_r(\xi)}\;,\label{eq:LerayUz}\\
   \xi {\mathcal P}'_z(\xi)&=&{S_\phi }(\xi)\;, \label{eq:LerayUr}\\
   \frac{1}{2}\xi S'_\phi(\xi)&=&-\frac{ U_r(\xi)+U_z(\xi)+ 1}{\nu +U_r(\xi)}  \, S_\phi(\xi)\;,\label{eq:LeraySphi}
\end{eqnarray}
where the function $S_\phi(\xi) $ in (\ref{eq:LerayUr}) and (\ref{eq:LeraySphi}) is related to the swirl velocity through 
\begin{eqnarray}
   S_\phi (\xi)=U_\phi(\xi)^2\geq 0\;.\label{eq:SphiDef}
\end{eqnarray}
Finally, we notice that  the dynamical system (\ref{eq:LerayDiv},\ref{eq:LerayUz},\ref{eq:LerayUr},\ref{eq:LeraySphi}) possesses a scale invariance: $\xi \to \lambda \xi $, which results from the absence of intrinsic physical scale in Euler equations. This scale invariance allows us to fix a convenient scale for the solution.

According with the boundary conditions (\ref{eq:BC1}) and (\ref{eq:BC2}), the dynamical system  is subjected to the boundary conditions
\begin{eqnarray}
S_\phi(0)&=&0\;, \label{eq:BCSphi}\\
U_r(0)&\neq &0\;, \label{eq:BCUr}\\
U_z(0)\neq 0 \quad &\& &  \quad  U'_z(0)= 0\;, \label{eq:BCUzi}\\
{\mathcal P}_z(0)&\neq&0\;.\label{eq:BCP1}
\end{eqnarray}
Moreover, since $U_\phi = \sqrt{S_\phi}$, the assumption of regularity of $U_\phi$ at $\xi=0$ requires that
\begin{eqnarray}
S'_\phi(0)&=&0\;, \label{eq:BCSphiprime}
\end{eqnarray}
which requires that $S_\phi(\xi)$ behaves at least as $S_\phi \sim \xi^2$ for $\xi\rightarrow0$. 
   
In the following, we study the generic behaviour of the dynamical system~ (\ref{eq:LerayDiv},\ref{eq:LerayUz},\ref{eq:LerayUr},\ref{eq:LeraySphi}) with the boundary conditions (\ref{eq:BCSphi},\ref{eq:BCUr},\ref{eq:BCUzi},\ref{eq:BCP1}).
 
\section{The solutions of the dynamical system}
\label{Secc:quatre}

A direct inspection of the dynamical system~(\ref{eq:LerayDiv},\ref{eq:LerayUz},\ref{eq:LerayUr},\ref{eq:LeraySphi})  shows that there exists a family of constant solutions given by
\begin{eqnarray}
   U_r(\xi)&=&a_0\;,\label{eq:BCUr0}\\
   U_z(\xi)&=&-2 a_0\;,\label{eq:BCUz0}\\
   {\mathcal P_z}(\xi)&=&p_0=a_0-2a_0^2\;,\label{eq:BCP0}\\
   S_\phi(\xi)&=&0\;,\label{eq:BCS0}
\end{eqnarray}
where $a_0$ is an arbitrary parameter. In the following, we seek other solutions satisfying regularity conditions along the $z$-axis by expressing the functions of the dynamical system as asymptotic series in the neighbhourhood of $\xi=0$.

 \subsection{Asymptotic expansions near $\xi=0$}

Let us start by the Taylor expansions of the functions $U_r(\xi)$ and $S_\phi(\xi)$ as given by
\begin{eqnarray}
   U_r(\xi)&=&\sum_{n=0}^\infty a_n \xi^{n} \;,\label{eq:AsymptoticSeriesUr}\\
   S_\phi(\xi)&=&\sum_{n=0}^\infty c_n \xi^{n}\;, \label{eq:AsymptoticSeriesS}
\end{eqnarray}
with real coefficients $a_n$ and $c_n$. Then, using Equations~(\ref{eq:LerayDiv},\ref{eq:LerayUr}), the asymptotic series representations of the functions $U_z(\xi)$ and ${\mathcal P_z}(\xi)$ are given by
\begin{eqnarray}
      U_z(\xi)&=&-\sum_{n=0}^\infty (2+n)a_n \xi^{n}\; ,\label{eq:AsymptoticSeriesUz}\\
   {\mathcal P_z}(\xi)&=&p_0 +  \sum_{n=1}^\infty \frac{c_n}{n} \xi^{n}\;,\label{eq:AsymptoticSeriesPz}
\end{eqnarray}
where $p_0$ is a real constant. Introducing the series representations~(\ref{eq:AsymptoticSeriesUr},\ref{eq:AsymptoticSeriesS},\ref{eq:AsymptoticSeriesUz},\ref{eq:AsymptoticSeriesPz}) into Equations~(\ref{eq:LerayUz},\ref{eq:LeraySphi}) and into the boundary conditions~(\ref{eq:BCSphi},\ref{eq:BCUzi}) allows us to determine the coefficients $p_0$, $a_n$ and $c_n$. 

First, the boundary conditions~(\ref{eq:BCSphi},\ref{eq:BCUzi}) require that
\begin{equation}
a_1=c_0=c_1=0\;,
\end{equation}
then, the leading order in the expansion of Eq.~(\ref{eq:LerayUz}) imposes that the constant $p_0$ is given by Eq.~(\ref{eq:BCP0}). The series expansions of  Equations~(\ref{eq:LerayUz},\ref{eq:LeraySphi}) up to fifth order lead to the following conditions
\begin{eqnarray}
c_2=0&\& &( 2 a_0-1 - 2\nu) a_2=0\;,\label{eq:order2}\\
 (a_0+2+ 3\nu ) c_3=0&\& &   15 (-1 - 3\nu + a_0) a_3 + 2 c_3 =0\;,\label{eq:order3}\\
 2 c_4 + 4\nu c_4 + 2 a_0 c_4=0&\& &  16 a_2^2 - 12 (1 + 4\nu) a_4 + c_4=0\;,\label{eq:order4}\\
- 3 a_2 c_3 + (2 + 5 \nu + 3 a_0) c_5=0&\& & 85 a_2 a_3 - 35 (1 + 5 \nu + a_0) a_5 + 2 c_5 =0\;,\label{eq:order5}
\end{eqnarray}
where we have used the condition $c_2=0$, given in Eq.~(\ref{eq:order2}), into the other equations to simplify their expressions. The second equation in (\ref{eq:order2}) yields two possibilities: (i) $a_0 =(1+2\nu)/2$, or (ii) $ a_2 =0$. The selection of the subsequent coefficients may be pursued up to all orders. However, different solutions are selected by the two branches (i) and (ii) defined above. In the following we study each branch separately.

\subsubsection{Case (i): A solution without swirl}
\label{Secc:SwirlessSolution}

Setting
\begin{equation}
a_0=\frac{1}{2}(1 + 2\nu)\;,
\end{equation}
in (\ref{eq:order2}) and simplifying Equations~(\ref{eq:order3},\ref{eq:order4},\ref{eq:order5}) yields
\begin{eqnarray}
 c_3=0&\& &    a_3 =0\;,\\
c_4=0&\& &16 a_2^2 - 12 (1 + 4\nu) a_4=0\;,\\
 c_5  =0&\& &  a_5=0\;.
\end{eqnarray}
By computing higher order terms in the series expansions of  Equations~(\ref{eq:LerayUz},\ref{eq:LeraySphi}) we show that $c_n=0$ for all $n\geq0$, $a_n=0$ for odd $n$ and $a_n\neq0$ for even $n$. Therefore, the solution of the branch corresponding to the first case satisfies $S_\phi(\xi)=U_\phi(\xi)=0$. In the following we will call it the {\it swirless} solution.

 Computing higher order terms of the series expansions of  Eq.~(\ref{eq:LerayUz}), one can show that the coefficients $a_{2n}$ for $n>1$ depend on the single coefficient $a_2$ and are explicitly given by
\begin{equation}
a_{2n}=\frac{1+4\nu}{2}\left(\frac{4a_2}{1+4\nu}\right)^{n-1}\frac{1}{n!}\quad \mathrm{for}\;n>1\;.
\end{equation}
Therefore, the series can be analytically summed up, yielding
\begin{equation}
 U_r(\xi)=  -\nu +  \left(1 + 4 \nu \right)\frac{1- e^{-\sigma\xi^2}}{2\sigma\xi^2}\;,
\end{equation}
where $\sigma=-4a_2/(1 + 4 \nu) $ and $\sigma>0$ to avoid divergent solutions at $\xi\rightarrow\infty$. Using the property of invariance of the system under the transformation $\xi\rightarrow\sqrt{\sigma}\xi$ one can take $\sigma=1$ without loss of generality. Therefore, the complete solution of this {\it swirless} flow is given by
\begin{eqnarray}
     {\mathcal P_z}(\xi) = -\nu (1 + 2 \nu) \quad & \& &  \quad 
    S_\phi(\xi) = 0\;,  \\
   U_r(\xi)=  -\nu +  \left(1 + 4 \nu \right)\frac{1- e^{-\xi^2}}{2\xi^2} \quad & \& &  \quad 
    U_z(\xi)= 2 \nu  -  \left(1 + 4 \nu \right) e^{-\xi^2}\;.\label{eq:UrUzZeroSwirl}
\end{eqnarray}
   
\begin{figure} 
\centerline{  \includegraphics[height=4.5cm]{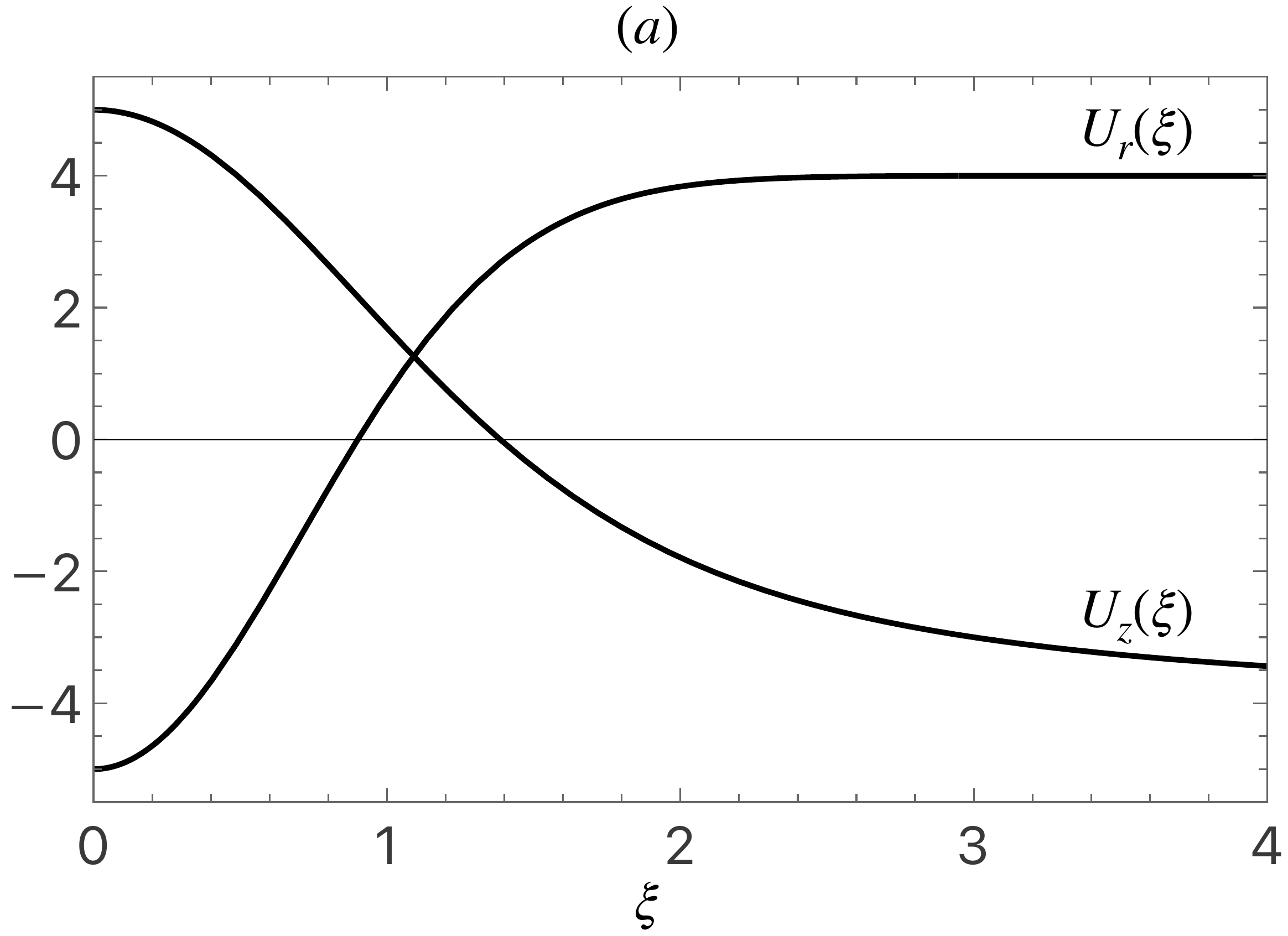}\includegraphics[height=4.5cm]{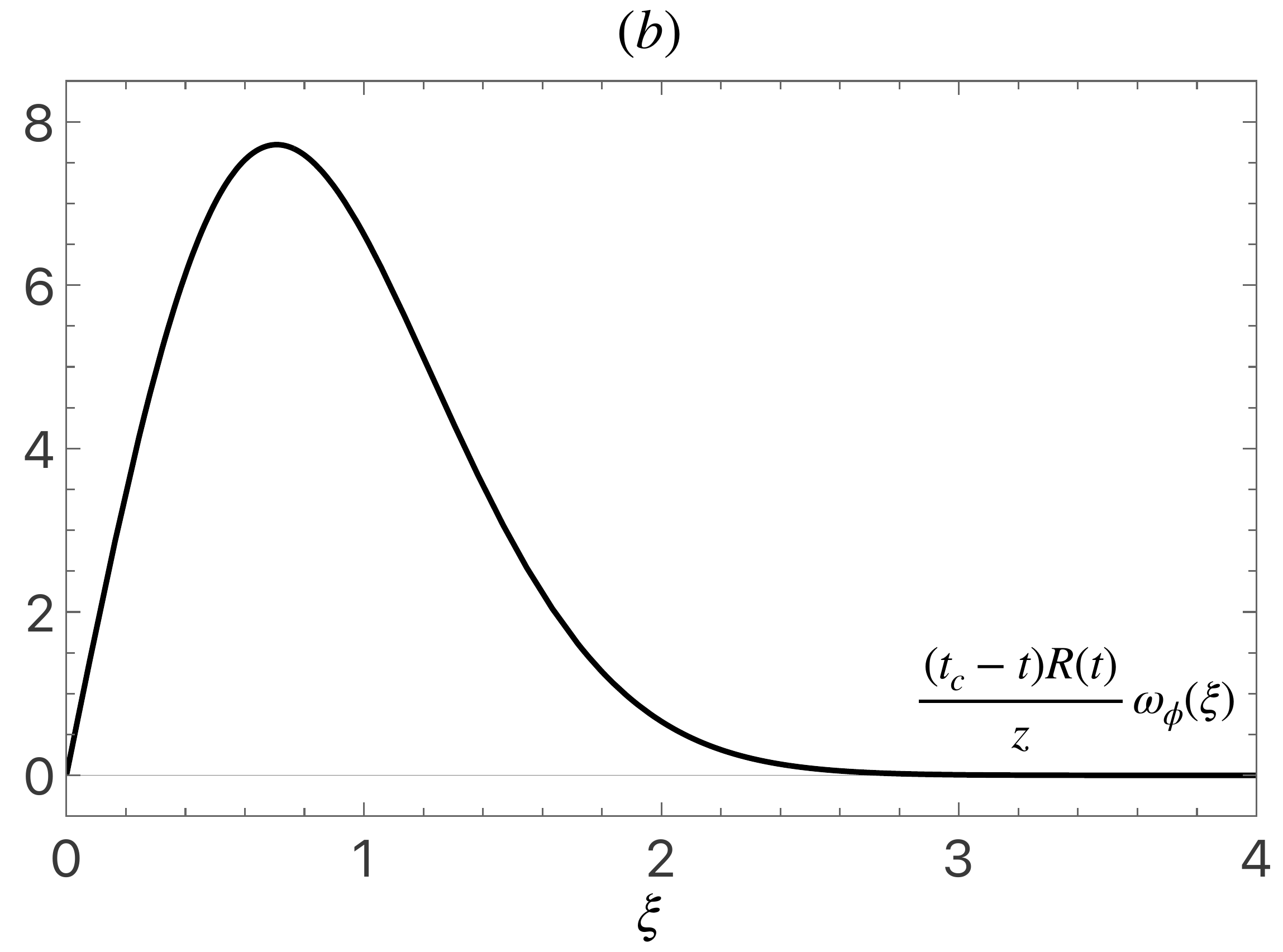} }
\caption{Solution with zero swirl for $\nu=2$. (a) $U_r(\xi)$  and $U_z(\xi)$ given by Eq.~(\ref{eq:UrUzZeroSwirl}). Notice that $U_r(\xi)$ and $U_z(\xi)$ vanish at $\xi>1$ and $\xi<1$ respectively.  (b) The self-similar vorticity $\omega_\phi(\xi)$ given by Eq.~(\ref{eq:VorticitySwirless}).} 
\label{fig:swirless}
\end{figure}
 
It is important to notice that while the flow is {\it swirless}, it is still vortical as the $\phi$-component of the vorticity is a non vanishing function of $\xi$:
\begin{equation}
\omega_r =  0\;,\quad \omega_\phi =  \frac{2(1+4\nu)}{(t_c-t) } \,  \frac{z }{R(t) }  \, \xi e^{-\xi^2}\;, \quad \omega_z =  0 \;  .\label{eq:VorticitySwirless}
\end{equation}
 However, the solution satisfies a vanishing total helicity, as the helicity density ${\bm u}\cdot{\bm \omega} =0$. Fig.~\ref{fig:swirless} shows the generic behaviour of the functions $U_r(\xi)$, $U_z(\xi)$ and $\omega_\phi(\xi)$. 
 
\subsubsection{Case (ii): A solution with swirl}
 
Consider the other possibility,
\begin{equation}
a_2=0\;,
\end{equation}
in (\ref{eq:order2}) and simplifying Equations~(\ref{eq:order3},\ref{eq:order4},\ref{eq:order5}) yields the following conditions
\begin{eqnarray}
 (a_0+2+ 3\nu ) c_3=0&\& &   15 (-1 - 3\nu + a_0) a_3 + 2 c_3 =0,\\
( 1 + 2\nu +  a_0) c_4=0&\& &  - 12 (1 + 4\nu) a_4 + c_4=0 , \label{eqx3} \\
 (2 + 5 \nu + 3 a_0) c_5=0&\& & - 35 (1 + 5 \nu + a_0) a_5 + 2 c_5 =0. 
 \end{eqnarray}
From the conditions~(\ref{eqx3}) and using higher order terms of the Taylor expansion, one can show that the first non trivial solution satisfies
\begin{eqnarray}
&&a_3=a_5=c_3=c_5=0\;,\\
&&a_0=-1-2\nu\quad\&\quad c_4=12(1+4\nu)a_4\;,\\
&&p_0= -(1+2 \nu ) (3+4 \nu)\;.
\end{eqnarray}
Here $a_4\neq 0$ is a free parameter. Therefore, $c_4\neq0$ and thus we uncover a solution with swirl that we will call a {\it swirled} flow. As one continues with the procedure, we can show that $a_p=c_p=0$ for all $p\neq 4n$. The first non zero coefficients of the series are shown in Table~\ref{Table:SolutionAsymptotic}. The coefficients $a_{4n}$ and $c_{4n}$ can be computed formally to any order $n$ using Mathematica$^{\mbox{\scriptsize{\textregistered}}}$. Unfortunately, to our knowledge, the series expansions for $U_r(\xi)$ and $S_\phi(\xi)$ cannot be expressed in terms of known analytic functions.

\begin{table}
\def~{\hphantom{0}}
\begin{center}
\begin{tabular}{ccccccccc}
$n$ & $0$ & $4$ & $8$ & $12$ &  $16$ & \dots\\ \hline
$a_n$ & $-(1+2 \nu)$ & $a_4$ & $\frac{ (4 \nu-5)}{20 (\nu +1)}a_4^2 $ & $\frac{(95-8 \nu 
   (2 \nu +25))}{560 (\nu +1)^2 (4 \nu +7)}a_4^3 $ & $ -\frac{  (8 \nu  (2
   \nu  (224 \nu  (\nu +2)+477)-437)+2635)}{20160 (\nu +1)^3 (4 \nu +7) (8
   \nu +11)} a_4^4$ &...\\ \hline
$c_n$ &  $0$ & $12 a_4 (4 \nu +1)$ & $ -\frac{18  (4 \nu +1)}{\nu +1}a_4^2$ & $ -\frac{3  (4 \nu +1)
   (28 \nu -65)}{20 (\nu  +1)^2}a_4^3 $ &$\frac{ (4 \nu +1) (4
   \nu  (344 \nu +541)-2605)}{140 (\nu +1)^3 (4 \nu +7)} a_4^4$ & ...\\
   \end{tabular}
   \caption{First five non zero coefficients $a_n$ and $c_n$ of the {\it swirled} solution.}
\label{Table:SolutionAsymptotic}
\end{center}
\end{table}

 Finally, it is important to notice that there exists infinite other possible solutions depending on the selection of the value of $a_0$, however the remaining solutions are series expansions in powers higher than $\xi^4$. In the following, we omit these solutions and focus on studying the properties of the {\it swirled} solution detailed in the following section.

\section{Properties of the {\it swirled} solution}\label{Secc:Swirled}

The asymptotic series of the {\it swirled} solution described previously are functions of the variable $a_4\xi^4$. To allow for numerical computation we use the truncated series
\begin{equation}
  U^{(N)}_r(\xi)=\sum_{n=0}^N a_{4n} \xi^{4n} \quad \& \quad
   S^{(N)}_\phi(\xi)=\sum_{n=0}^N c_{4n} \xi^{4n}\;, \label{eq:AsymptoticSeriesUrS4}
   \end{equation}
with $U^{(N)}_r(\xi)\rightarrow {U}_r(\xi)$ and $S_\phi^{(N)}(\xi)\rightarrow{S}_\phi(\xi)$ for $N\rightarrow\infty$. By varying the order of the expansion $N$ we obtain a convergent scheme for the solutions in terms of the asymptotic series (\ref{eq:AsymptoticSeriesUr},\ref{eq:AsymptoticSeriesS},\ref{eq:AsymptoticSeriesUz},\ref{eq:AsymptoticSeriesPz}). In addition, using the property of invariance of the system under the rescaling of the coordinate  $\xi$, the computation is highly simplified by setting $a_4=1$. Notice that we have assumed {\it a priori} that $a_4>0$; it can be shown that taking $a_4<0$ yields to unphysical divergent solutions.

\begin{figure} 
\centerline{\includegraphics[height=4.5cm]{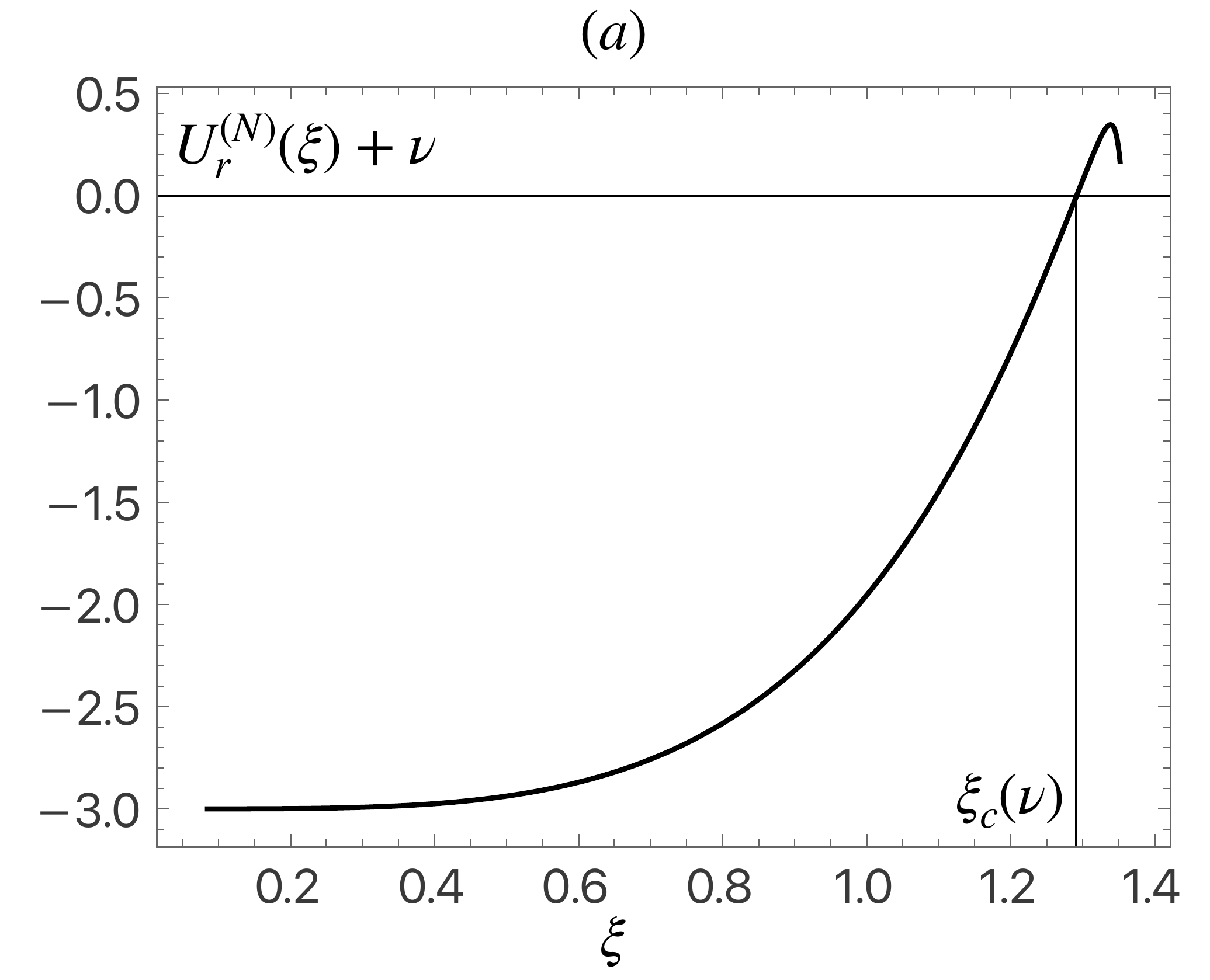}\includegraphics[height=4.5cm]{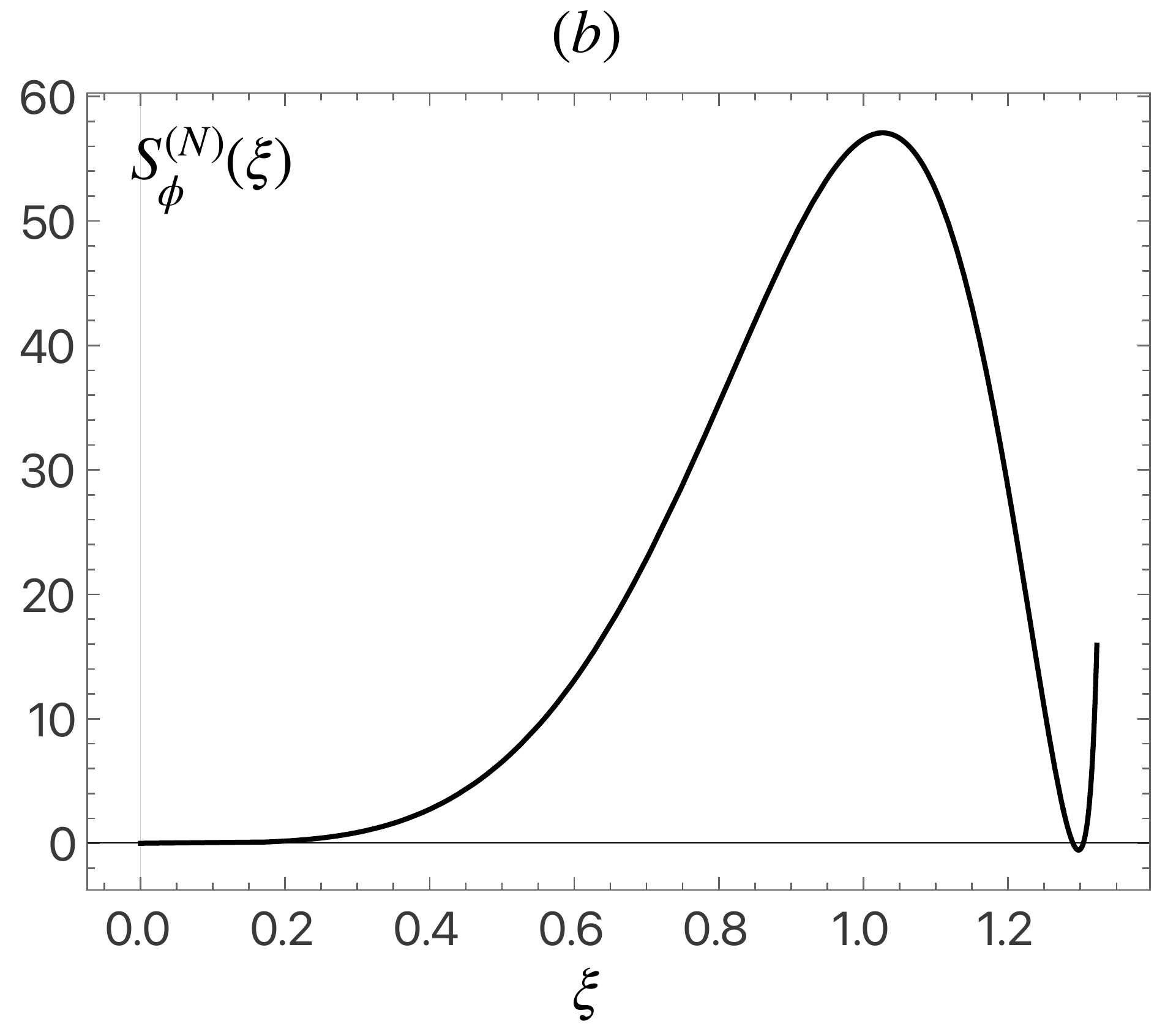} }\centerline{\includegraphics[height=4.5cm]{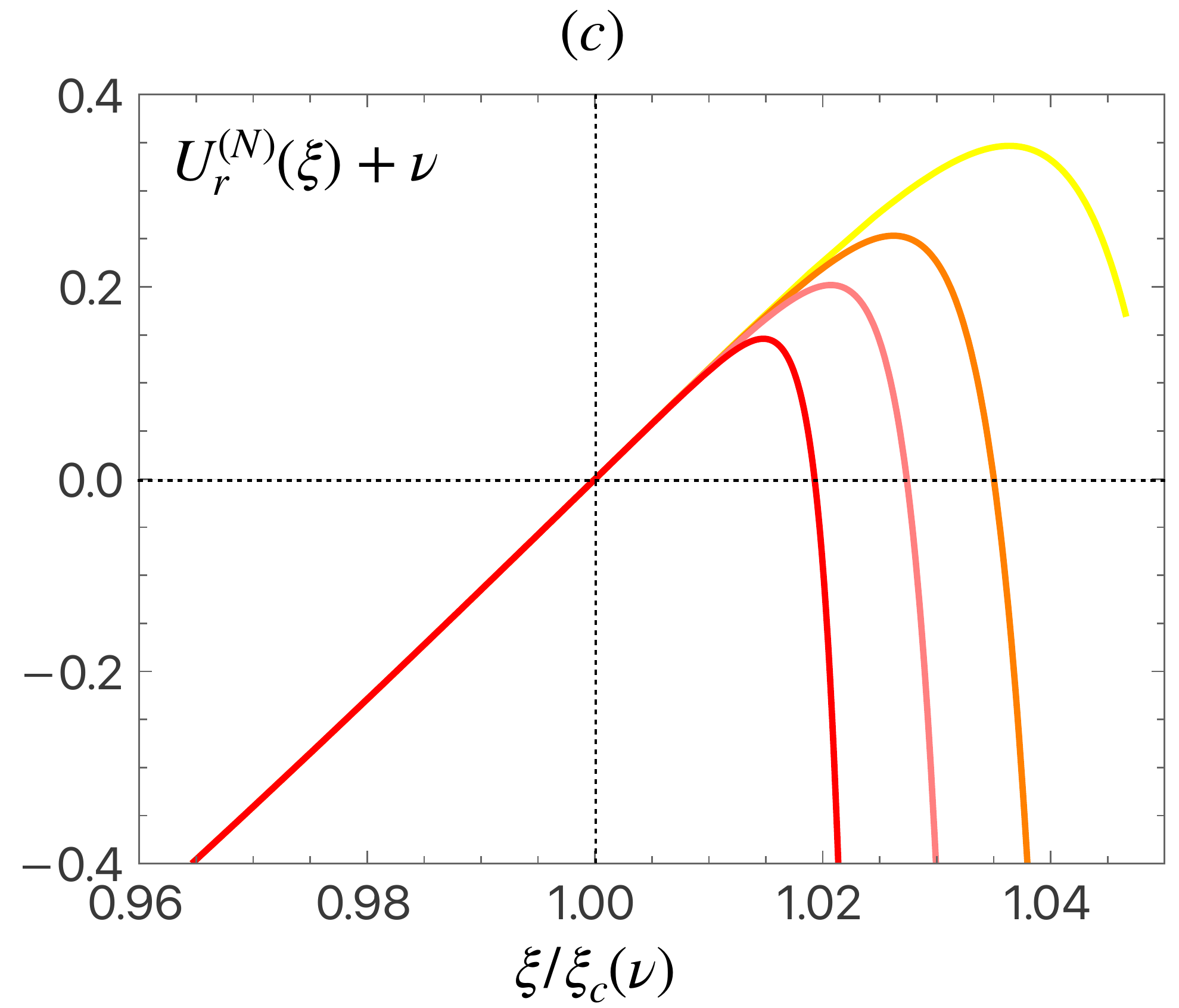} \includegraphics[height=4.5cm]{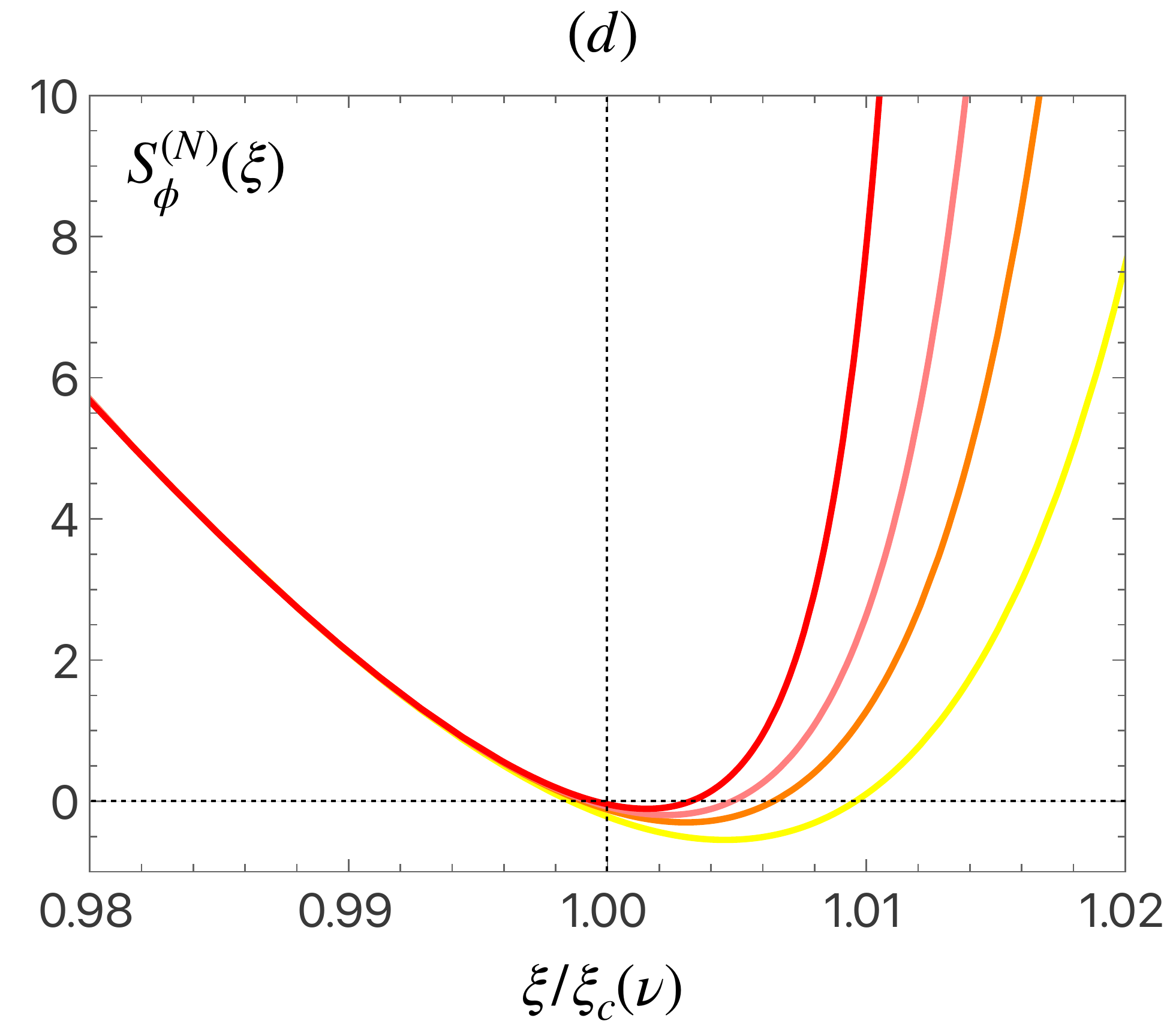} }
\caption{ \label{Fig:asymptotic} Plot of the series (a) $U^{(N)}_r(\xi)+\nu$ and (b) $S^{(N)}(\xi)$ for $N=50$ and $\nu=2$. The critical point $\xi_c\approx 1.29167$ is a solution of $U^{(N)}_r(\xi)+\nu=0$. (c) and (d)  Zooms on the functions $U^{(N)}_r(\xi/\xi_c)+\nu$ and $S^{(N)}(\xi/\xi_c)$ in the neighbourhood of $\xi=\xi_c(\nu)$. The different curves correspond to different values of the series order $N$ included in the expansion: $N=50$ (yellow), $N=75$  (orange),  $N=100$  (pink) and $N=150$  (red). The convergence of the series are reached at $\xi=\xi_c(\nu)$ satisfying $U_r(\xi_c)+\nu=S(\xi_c)=S'(\xi_c)=0$. } 
\end{figure}

Fig.~\ref{Fig:asymptotic}(a,b) shows an example of the generic behaviour of $U^{(N)}_r(\xi)$ and $S_\phi^{(N)}(\xi)$. The function $U^{(N)}_r(\xi)$ starts at $U^{(N)}_r(0)=-(1+2\nu)<0$ and increases monotonically with $\xi$ until it reaches $U^{(N)}_r(\xi)\approx -\nu$ at $\xi=\xi_c(\nu)$. At this point, the function $S_\phi^{(N)}(\xi_c)\to 0$. Indeed, Fig.~\ref{Fig:asymptotic}(c,d) shows that at the critical point $\xi=\xi_c(\nu)$, the conditions $U_r(\xi_c)+\nu\equiv 0$ and $S_\phi(\xi_c)\equiv 0$ are simultaneously satisfied for $N\rightarrow \infty$. Moreover, the extra condition~(\ref{eq:BCSphiprime}): $S'(\xi_c(\nu))=0$ is also satisfied. These results confirm that the radius of convergence of the asymptotic series (\ref{eq:AsymptoticSeriesUr},\ref{eq:AsymptoticSeriesS},\ref{eq:AsymptoticSeriesUz},\ref{eq:AsymptoticSeriesPz}) is indeed $\xi_c(\nu)$.
 
As a conclusion, the exploration of the possible solutions for different values of $\nu$ indicates that the dynamical system~(\ref{eq:LerayDiv},\ref{eq:LerayUz},\ref{eq:LerayUr},\ref{eq:LeraySphi}) has a critical point characterised by $U_r(\xi_c) = -\nu$. At this critical point, the swirl velocity $S_\phi(\xi_c)$ vanishes and may change its sign which is not allowed by the definition (\ref{eq:SphiDef}). This means that the solution of the dynamical system is physically relevant only for $\xi\leq \xi_c$ that should be matched with a different solution for $\xi>\xi_c$. In what follows we characterise the behaviour of the solution near the critical point.  

\begin{figure} 
\centerline{  \includegraphics[height=4.5cm]{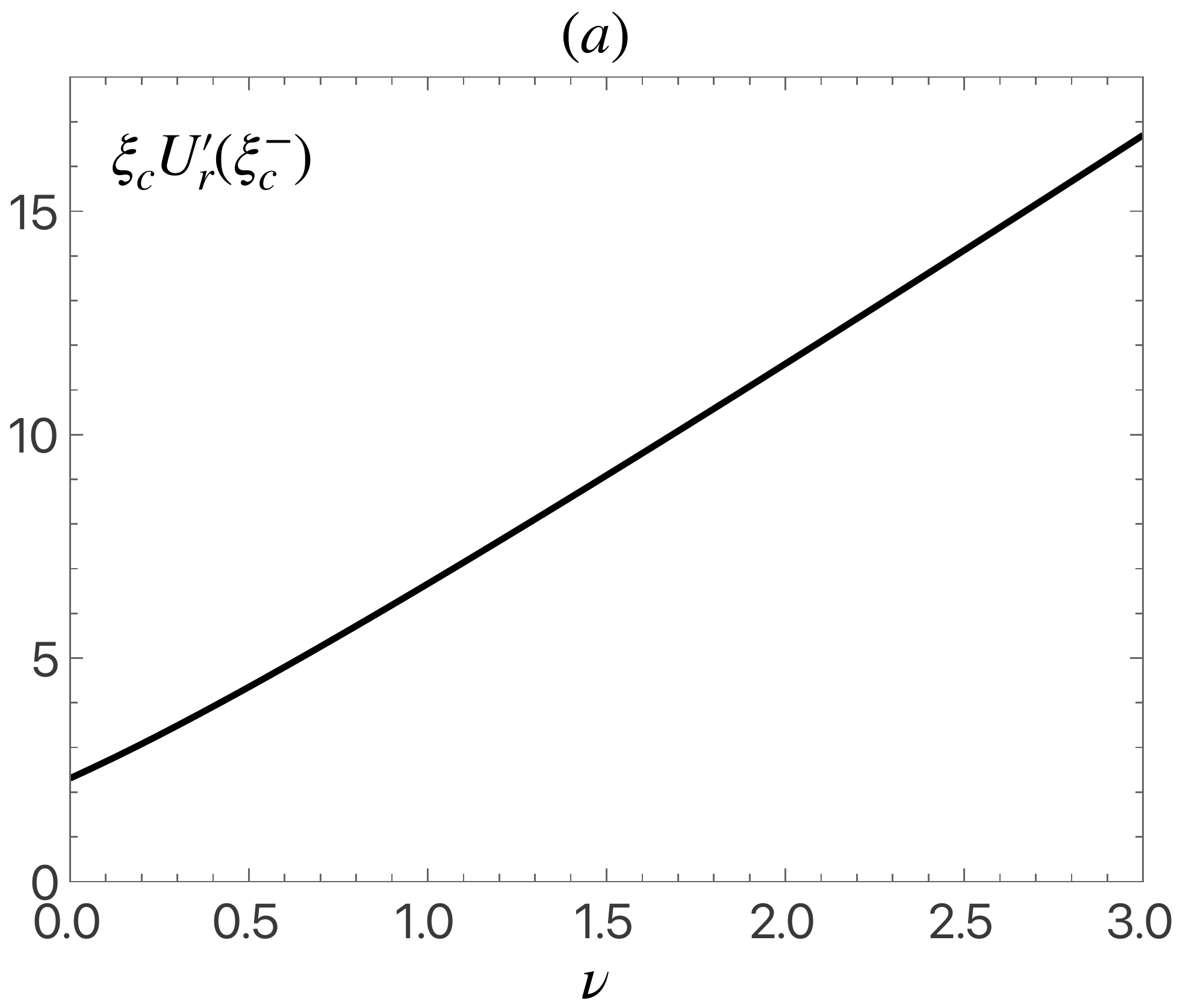}\includegraphics[height=4.5cm]{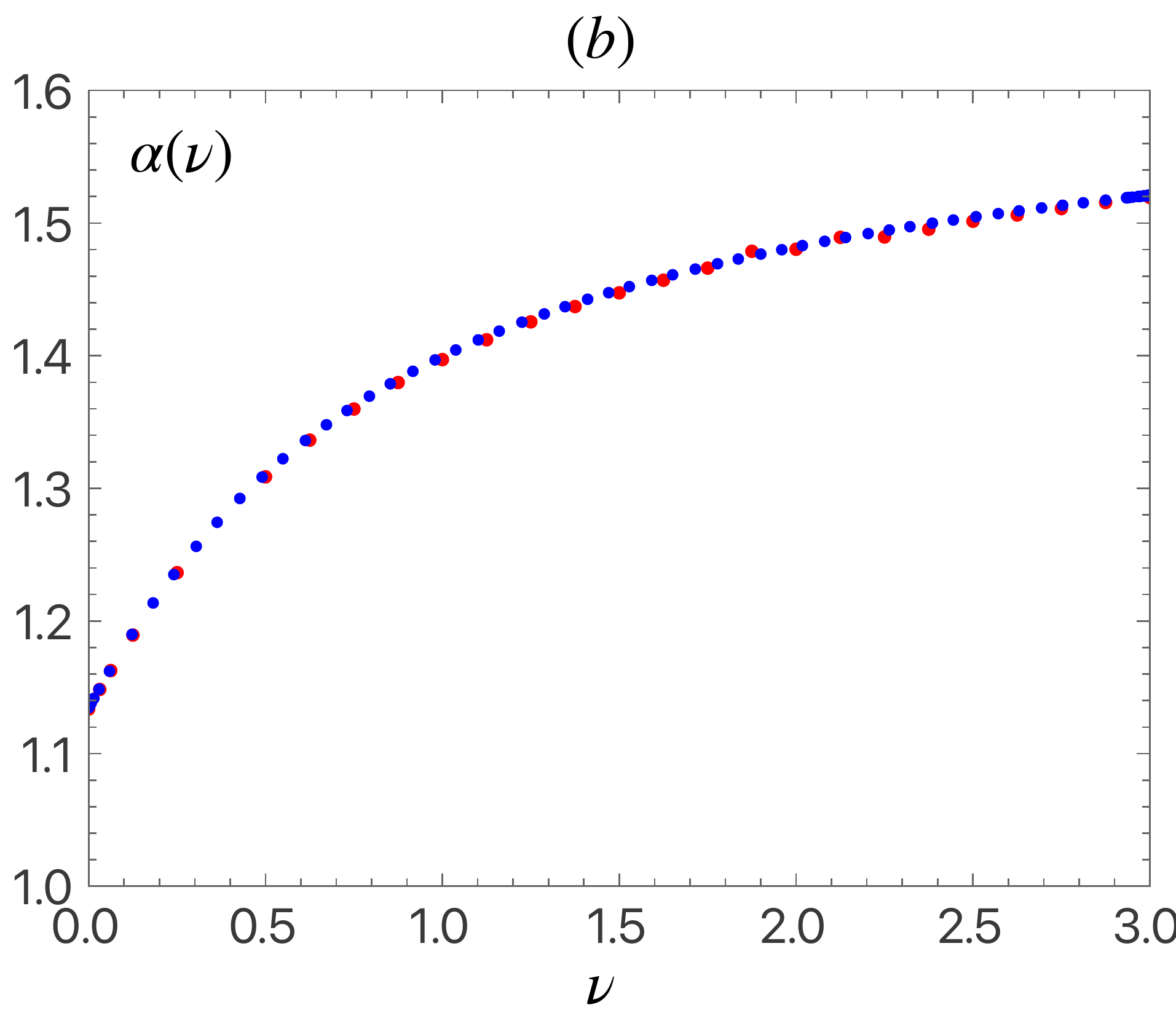}}
\caption{ \label{Fig:alphavsnu} (a) Numerical curve for the value of $\xi_c U'_r(\xi_c^-)$ as function of $\nu$. (b) Numerical curve for the value of the exponent $\alpha$ as a function of  $\nu$. Blue points are obtained by the interpolation of the asymptotic series. Red dots represents the numerical  solution of the ordinary differential equations with the right boundary conditions.} 
\end{figure}

Expanding
\begin{equation}
U_r(\xi)\approx -\nu - U'_r(\xi_c^-) (\xi_c-\xi)+ \dots\;,
\end{equation}
where the dependance of $U'_r(\xi_c^-)$ as function of $\nu$ is shown in Fig.~\ref{Fig:alphavsnu}(a). Then using equation (\ref{eq:LerayDiv}) one obtains up to first order:
\begin{equation}
U_z(\xi)\approx 2 \nu- \xi_c  U'_r(\xi_c^-) + \dots
\end{equation}
Therefore, Equation~(\ref{eq:LeraySphi}) implies 
\begin{equation}
-\frac{1}{2} \xi_c U'_r(\xi_c^-)  (\xi_c-\xi) S'_\phi(\xi)\approx- (1+\nu - \xi_c  U'_r(\xi_c^-))
   \, S_\phi(\xi)\;,
\end{equation}
that is,
\begin{eqnarray}
S_\phi(\xi)\sim (\xi_c -\xi)^{\alpha(\nu)}, \quad {\rm with} \quad \alpha(\nu) =  2 -\frac{2 (1+\nu )}{\xi_c  U'_r(\xi_c^-)}\;  .\label{eq:LeraySphiNearxic} 
   \end{eqnarray}
Notice that because $U'(\xi) >0$, Eq.~(\ref{eq:LeraySphiNearxic}) imposes $\alpha(\nu) < 2$. Moreover, the condition $S'(\xi_c)=0$ imposes $\alpha(\nu)>1$. On the other hand, as a consequence  of the  invariance of the dynamical system~(\ref{eq:LerayDiv},\ref{eq:LerayUz},\ref{eq:LerayUr},\ref{eq:LeraySphi}) under the transformation $\xi\to\lambda\xi$, the value of $\alpha$ does not depend explicitly on $\xi_c$. These properties are confirmed when $\alpha$ is  computed for various values of $\nu$, through Eq.~(\ref{eq:LeraySphiNearxic}) and the use of the series expansion up some high order $N$, that provides a confident estimate (see Fig.~\ref{Fig:alphavsnu}(b)). Finally, the explicit dependence of $\xi_c(\nu)$ is not relevant because of the scale invariance of the dynamical system.
   
\subsection{Direct numerical resolution}

In addition, we solve numerically the dynamical system~(\ref{eq:LerayDiv},\ref{eq:LerayUz},\ref{eq:LerayUr},\ref{eq:LeraySphi}) with initial boundary conditions (\ref{eq:BCSphi},\ref{eq:BCUr},\ref{eq:BCUzi},\ref{eq:BCP1}) by usual integration algorithms. The result of the numerical solutions of the ordinary differential equations  agree with a relative error for the functions of the order of $0.4 \%.$   As in previous section we observe that the self similar swirl $S_\phi(\xi) $ vanishes at some point, as well as, at the same point $U_r(\xi_c) \to  -\nu$. Finally, using the same formula (\ref{eq:LeraySphiNearxic}), the exponent $\alpha(\nu)$ is also estimated. Naturally, as shown in Fig.~\ref{Fig:alphavsnu}, the computation of the parameter $\alpha$ as a function of $\nu$ using high order asymptotic series and the traditional numerical solution of ordinary differential equations agree satisfactorily. 

 \subsection{Solution near the critical point}
 
In this section, we study in more detail the local expression for the solutions of the self-similar functions $S_\phi(\xi)$,  $U_r(\xi)$, $U_z(\xi)$ and  ${\mathcal P_z}(\xi) $ near the critical point. From now on, we will set $\xi_c(\nu)\equiv1$ by using, again, the property of scale invariance of Euler equations.

The calculations show that near the critical point the dynamical system  admits a solution of the form    
\begin{eqnarray}    S_\phi(\xi) 
   &\approx& \sigma_c  (1-\xi)^{\alpha }  + \dots \label{eq:OuterSphi} \\
   U_r(\xi)&\approx& -\nu+ \frac{a \left(1-\xi ^2\right)}{\xi ^2}-\frac{2 \sigma_c (1-\xi )^{\alpha +2}}{(\alpha +1) (\alpha +2) (2  a(\alpha-1)- 4 \nu -1)}+\dots   \label{eq:OuterUr}  \\
   U_z(\xi) 
   &\approx& 2 (a+\nu )-\frac{2 \sigma_c (1-\xi )^{\alpha +1}}{(\alpha +1) (2 a (\alpha -1)-4 \nu
   -1)}  +\dots    \label{eq:OuterUz}  \\
   {\mathcal P_z}(\xi) 
   &\approx&-(a+\nu ) (2 a+2 \nu +1)-\frac{\sigma _c (1-\xi )^{\alpha +1}}{\alpha +1} + \dots  . \label{eq:OuterPz} 
\end{eqnarray}
Notice that the consistence with Eq.~(\ref{eq:LeraySphiNearxic}) imposes 
\begin{equation}
 \alpha=2+ \frac{1+\nu }{a } \quad\&\quad U'_r(1^-) = - 2a >0\;.\label{eq:valueOfa}
\end{equation}

\subsection{Solution beyond the critical point $\xi>1$}

Outside the tubular region $\xi=1$, the swirl is identically zero
\begin{equation}
U_\phi(\xi>1)=0\;.
\label{eq:OuterUphi1}
\end{equation}
Setting $S_\phi(\xi)=0$ into the dynamical system~(\ref{eq:LerayDiv},\ref{eq:LerayUz},\ref{eq:LerayUr},\ref{eq:LeraySphi}), one obtains ${\mathcal P}'_z(\xi)=0$, thus by continuity of (\ref{eq:OuterPz}) one obtains
 \begin{equation}
{\mathcal P_z}(\xi)=-\left(a+\nu \right) \left(2(a+ \nu) +1\right)\;,
      \end{equation}
 for $\xi>1$, with $U'_r(1^-)=-2a$ (see (\ref{eq:valueOfa})). The equations for $U_r$ and $U_z$ become
     \begin{eqnarray}
 \xi U'_z(\xi ) &=&-\frac{\left( U_z(\xi) -2(a+\nu) \right)\left( U_z(\xi) +  2(a+\nu)+1\right) }{\nu
   +{U_r}(\xi)}\;,\label{eq:LerayUzBIS}\\
  \xi U'_r(\xi)&=&-2 {U_r}(\xi)-{U_z}(\xi)\;.\label{eq:LerayDivBIS}
   \end{eqnarray}
   The boundary conditions are given by imposing continuity of (\ref{eq:OuterUr})  and  (\ref{eq:OuterUz})
\begin{eqnarray}
{U_r}(1)=-\nu  \quad \& \quad    U_z(1) = 2(a+\nu)\;.\label{eq:StartingPoint}   \end{eqnarray} 
The relevant solution of this system yields
     \begin{eqnarray}
{U_r}(\xi)=-\nu -\frac{U_r'(1^-)}{2}\left(\frac{1-\xi ^2}{\xi ^2} \right)\quad\&\quad  {U_z}(\xi )= 2\nu-U_r'(1^-)\;. \label{eq:OuterUr1}
   \end{eqnarray}
This solution agrees with the leading terms in Eqns. (\ref{eq:OuterUr}) and (\ref{eq:OuterUz}). 
   
\begin{figure} 
\centerline{  \includegraphics[height=4.cm]{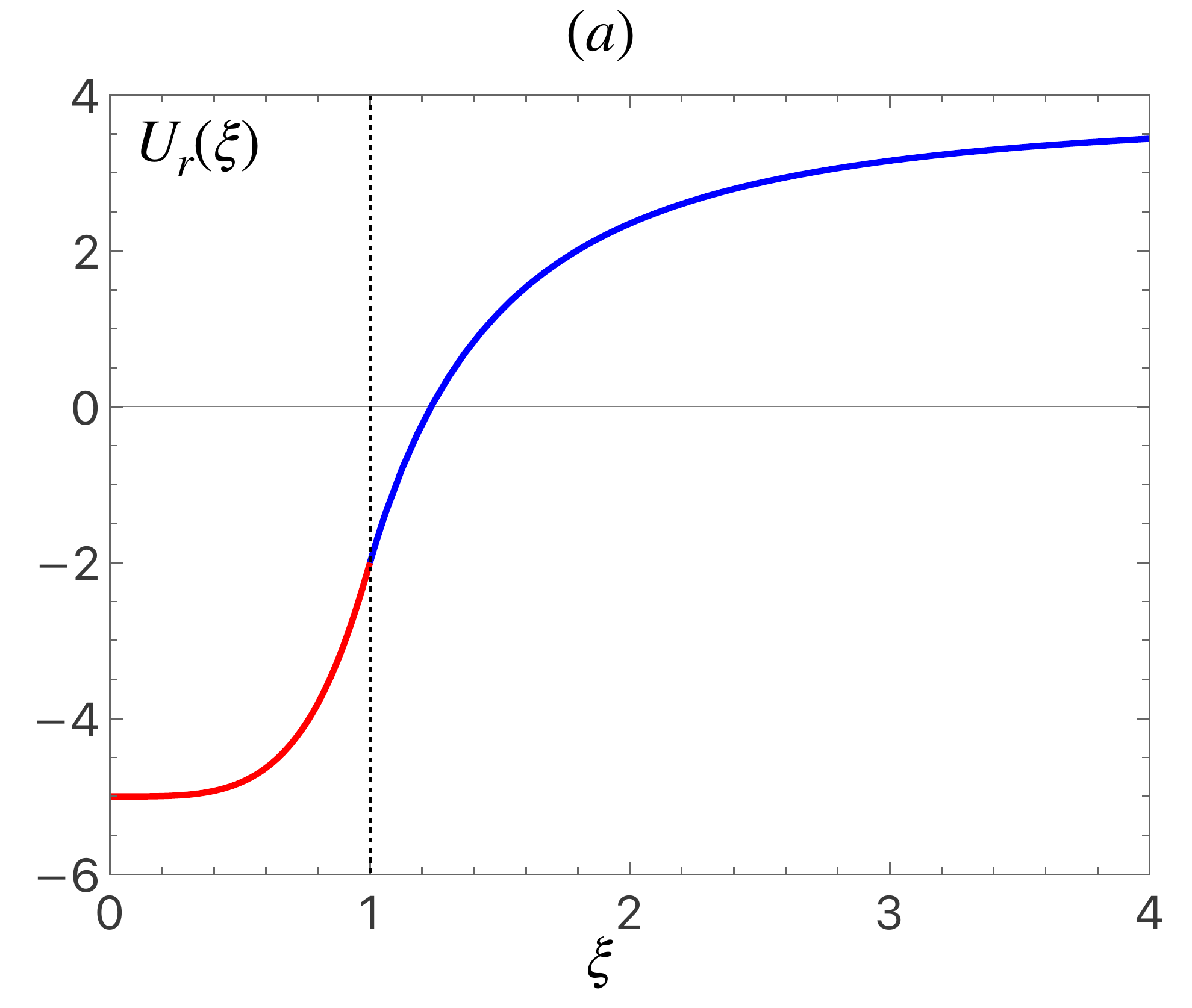}\includegraphics[height=4.cm]{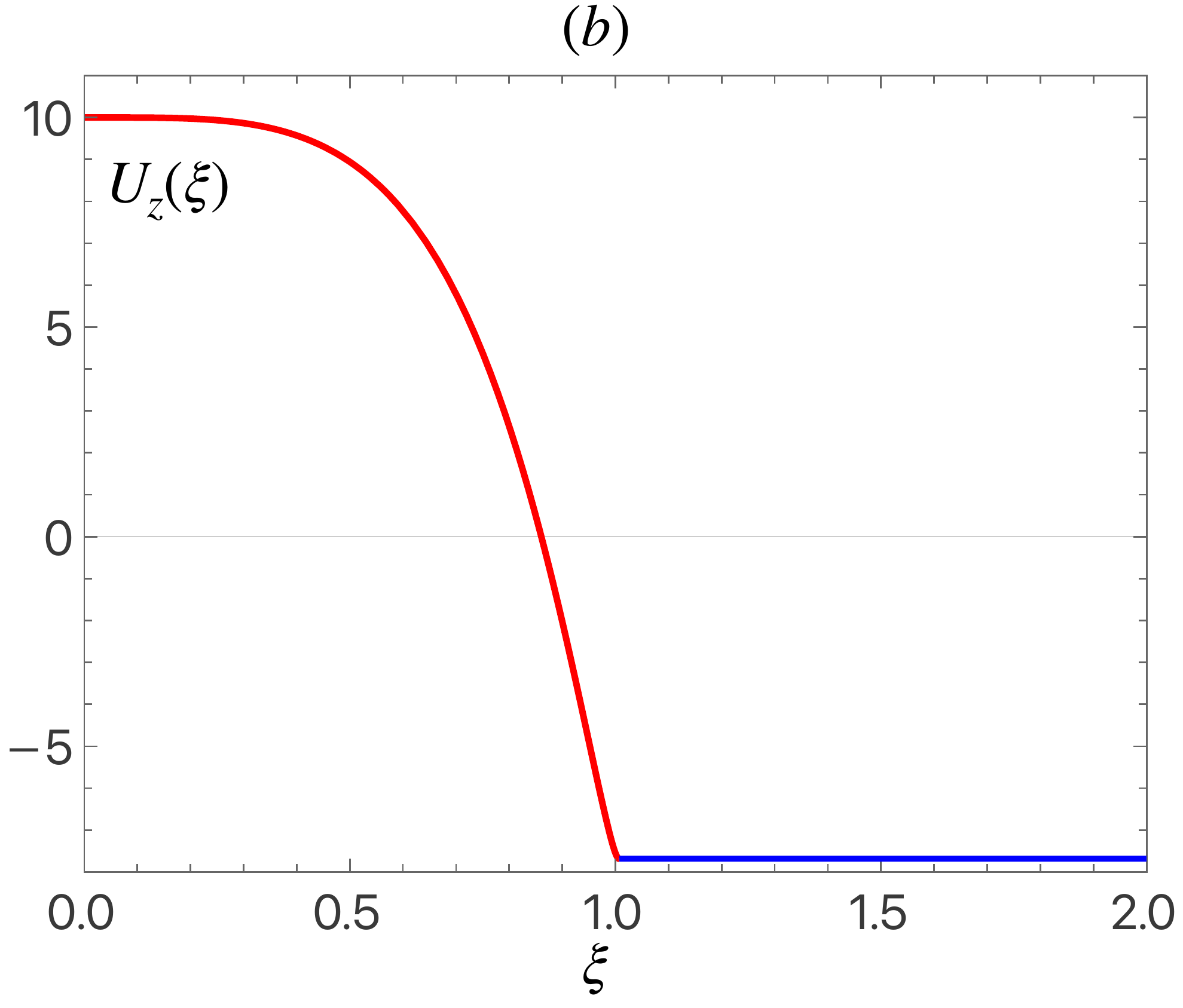}\includegraphics[height=4.cm]{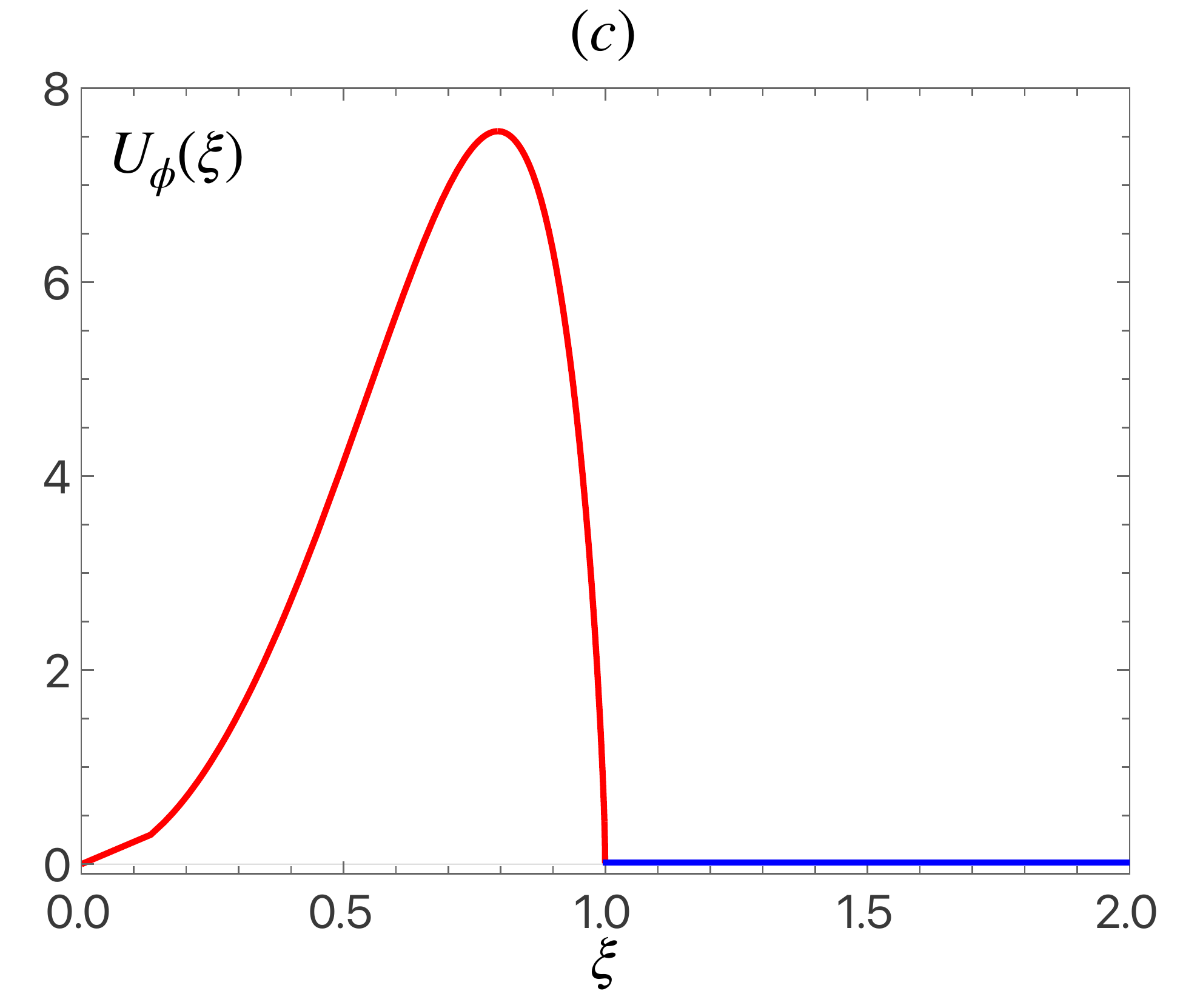}}
\caption{ \label{Fig:ODEsSolution} Numerical solution of the functions (a) $U_r(\xi)$, (b) $U_z(\xi)$ and (c) $U_\phi(\xi)$ for $\nu=2$. For $\xi<1$, the red curves show the asymptotic series (\ref{eq:AsymptoticSeriesUr},\ref{eq:AsymptoticSeriesS},\ref{eq:AsymptoticSeriesUz}) with $N=100$, that is up to an order ${\mathcal O}(\xi^{400})$. For $\xi>1$, the blue curves show the exact solutions~(\ref{eq:OuterUphi1},\ref{eq:OuterUr1}).}
\end{figure}
   
In Fig. \ref{Fig:ODEsSolution} we plot the self similar functions $U_r(\xi)$, $U_z(\xi)$, and $U_\phi(\xi)$,  for the special case $\nu=2$. The asymptotic series  defined by (\ref{eq:AsymptoticSeriesUrS4}) are drawn in red color. The solutions~(\ref{eq:OuterUr1}) are drawn in blue as a continuation for $\xi>1$. Finally, notice that the behaviour of the different functions near $\xi=1$ show that $U_r(\xi)$ is of class $C^3$, $U_z(\xi)$ and $P_z(\xi)$ are of $C^2$ and $S_\phi(\xi)$ of Class $C^1$. On the other hand, $U_\phi(\xi) = \sqrt{S_\phi(\xi)}$  is just a continuous function, thus a $C^0$ function.
  
 \section{ Behaviour of Conserved quantities prior to blow-up}\label{Secc:Conserved}
 
 \subsection{Sketch of the {\it swirled} solution in the real space}
 
 \begin{figure}
\centerline{\includegraphics[height=6cm]{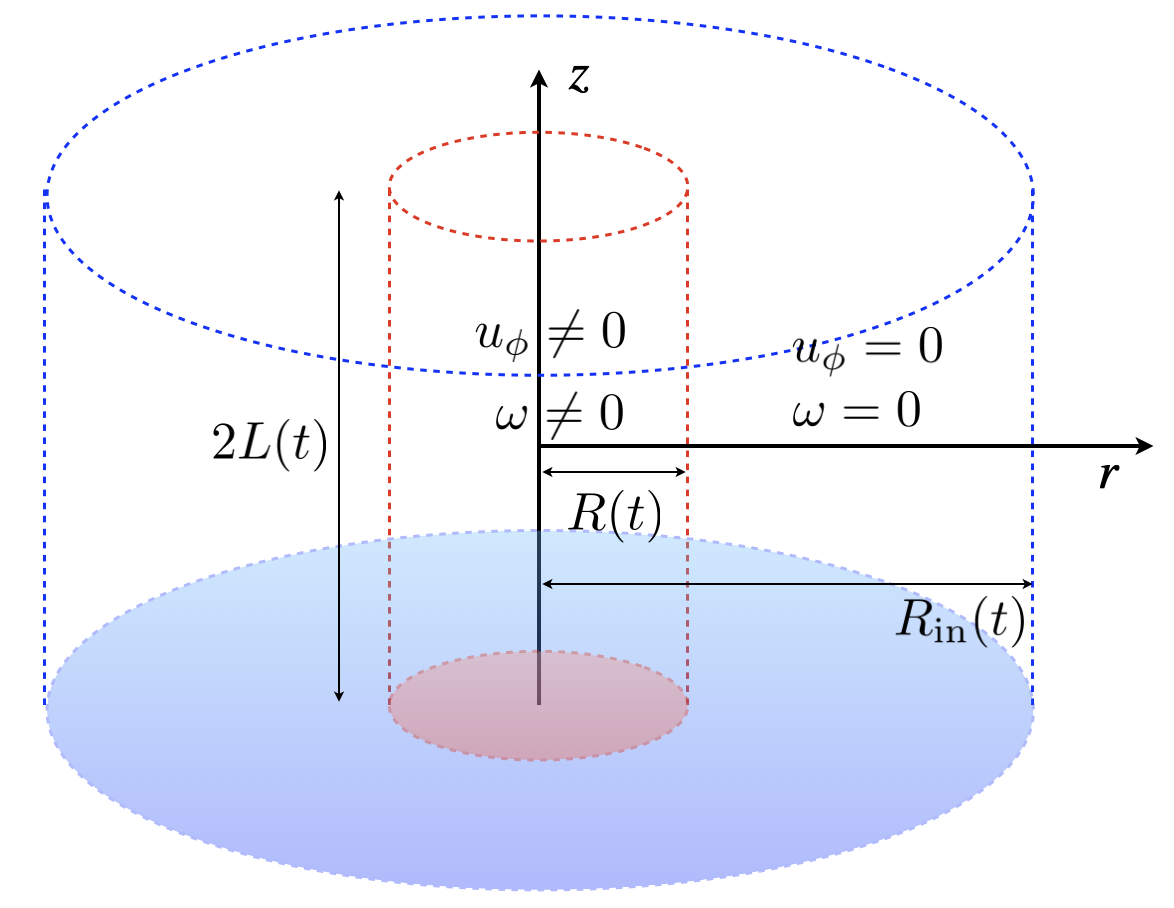}}
\caption{ \label{Fig:Scheme} Scheme for the complete solution of Euler equations. The whole space domain is divided into an inner and an outer region. The solution in the inner region is described by a self-similar solution and consists of a cylindrical region of a radius $R_{\rm in}(t)$ and a vertical length of $2L(t)$. The outer region ensures the right behaviour as the radial, $r$, and the vertical, $z$, distances tend to infinity. Inside the cylindrical region delimited by a radius $R(t)$ the swirl and vorticity are not zero, while outside this region vorticity and swirl vanish. The separation of variable leads to $ R(t)\sim (t_c-t)^\nu$ with $\nu$ a constant. For $\nu>0$ the tubular region shrinks either into a line or a point in finite time, depending on the behaviour of $L(t)$. While, the self-similar behaviour of the flow imposes $R_{\mathrm{in}}(t)=\xi_{\mathrm{in}}R(t)$ with $\xi_{\mathrm{in}}>1$, the dynamics of $L(t)$ should be derived from the theory.}
\end{figure}

Up to this point we have proven the existence of a finite-time singular flow that was alluded to in the Introduction. To be more precise, this self-similar solution is asymptotically valid within an inner cylindrical volume characterised by $\mathbb{V}_{\rm in}=[0,R_{\mathrm{in}}(t)]\times[-L(t),L(t)]$ (See Fig.~\ref{Fig:Scheme}). The self-similar behaviour is restricted to an inner solution that should  match  an outer solution which is valid within a complementary volume ${V}_{\rm out}=\mathbb{R}^3-\mathbb{V}_{\rm in}$ and ensures the boundary conditions at infinity in such a way that the complete solution satisfies the conserved quantities. The only condition imposed to the outer flow is that it remains {\it swirless} and with a zero vorticity. Indeed, the swirl remains zero because the Euler equation for the swirl component (\ref{eq:uphi}) may be written as $\frac{d}{dt }(r u_\phi)=0$, so that $r u_\phi $ remains constant (and zero) along characteristic curves. On the other hand, for $r>R(t)$ the potential flow cannot generate vorticity smoothly. Therefore, in the region $r> R(t)$ (or $1<\xi<\infty$) the full flow remains {\it swirless} and has zero vorticity. Finally, the inner solution (\ref{eq:OuterUr1}) should match at a region of the order of $R_{\mathrm{in}}(t)$ with an outer potential solution that ensures the proper boundary condition  for convergence. Since self-similar behaviour holds in the inner region, the temporal evolution of $R_{\mathrm{in}}(t)$ occurs on the same time scale as that of $R(t)$. That is, $ R_{\mathrm{in}}(t)/R(t) \sim \xi_{\mathrm{in}}$,
where  $\xi_{\mathrm{in}} > 1$ is a constant.~\footnote{An asymptotic matching works, as a general rule, if $\xi_{\mathrm{in}} \gg 1$. However the explicit determination of this region depends on the initial conditions.}
 
 \begin{figure} 
\centerline{\includegraphics[width=12cm]{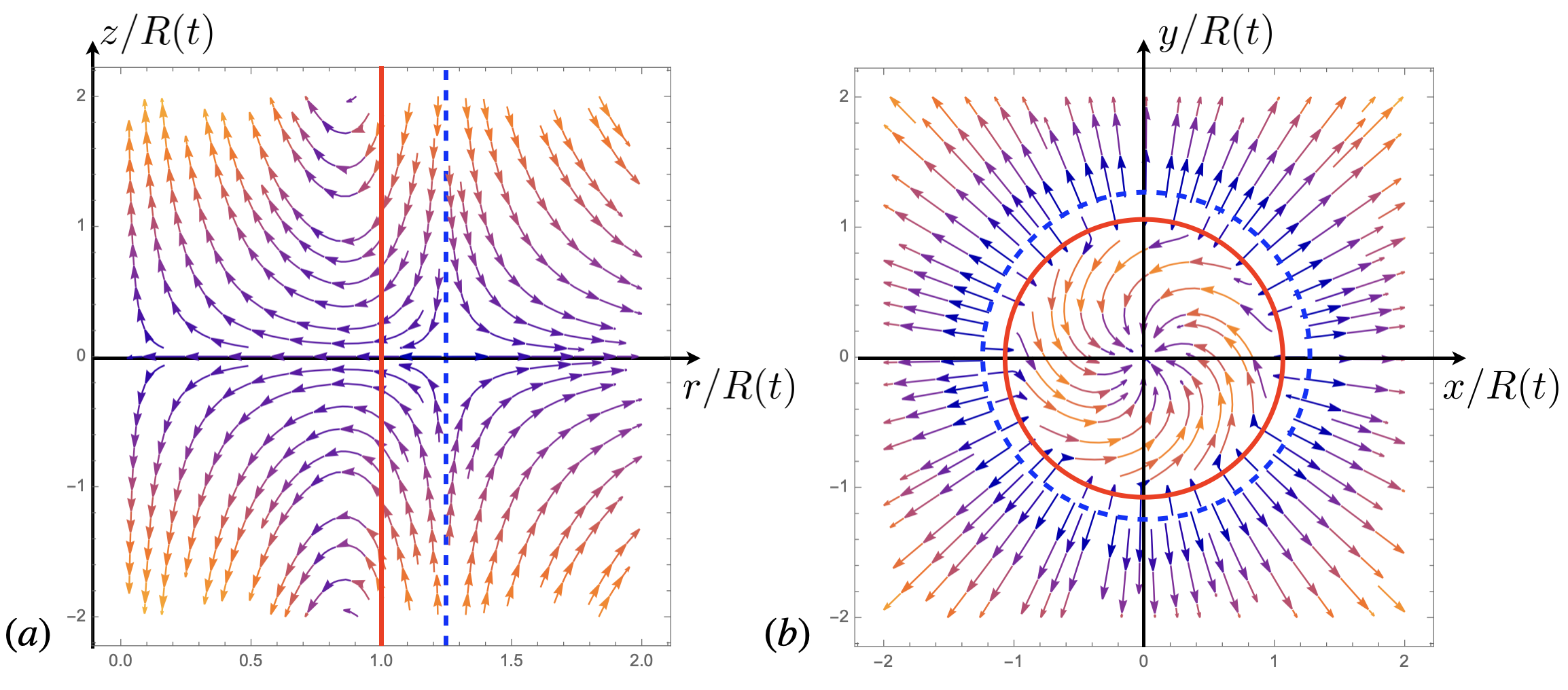}}
\caption{ \label{Fig:Stream2D}   The streamlines of the {\it swirled} case for $\nu=2$ corresponding to the flow of the scaled axisymmetric velocity field, ${(t_c-t)}\,{\bm u}(\bm{x},t)\, /{R(t)}$, as function of the dimensionless coordinates $(r/R(t),z/R(t))$. Panel (a) shows the streamlines in the $(r,z)$-plane, and, panel (b) shows streamlines in the $(x,y)$-plane at $z=R(t)$. The red solid line in (a) and the circle in (b) correspond to the surface $\xi=\xi_c=1$, for which $U_r(\xi_c)+\nu=0$, separating the {\it swirled} phase $(\xi<\xi_c=1)$ from the {\it swirless} phase $(\xi>\xi_c=1)$. The dotted blue line in (a) and the circle in (b) correspond to the surface $U_r(\xi_0)=0$ where the streamlines change direction from inward ($\xi<\xi_0$) to outward ($\xi>\xi_0$). Because $\nu>0$, one has that $\xi_0> \xi_c$. } 
\end{figure}

Finally, if one inspects the streamlines of the {\it swirled} solution in the inner region, the projection in $(r,z)$-plane shows that the flow in the neighborhood of $r=0$ is inward the $z$-axis conjecturing that the {\it swirled} case has a focusing mechanism that induces a blow-up. Along the $(x,y)$-plane the flow shows a structure (pattern) that is reminiscent of interfaces separating two-phases: here a {\it swirled} vortical region inside a tubular volume and a region without swirl nor vorticity. Along the radial direction, the tubular region shrinks as a power law $(t_c-t)^\nu$ vanishing at $t=t_c$.
 
\subsection{Kinetic energy and helicity of the {\it swirled} case}
 \label{sec:swirled}

In this section, we confront the self-similar solution with {\it all} conservations laws satisfied by inviscid flows obeying Euler equations \citep{majda_bertozzi_2001}. 

The conservation laws are be divided into two groups. The first group involves vectorial conservation laws associated with the total momentum, $\bm{V}$, the total vorticity, $\bm{\Omega}$, the fluid impulse, $\bm{I}$, and the moment of the fluid impulse, $\bm{M}$. They are explicitly given by~\citep{majda_bertozzi_2001}
 \begin{eqnarray}
 \bm{V}&=&\int_{\mathbb{R}^3} \bm{u}\,d^3x\;,\\
  \bm{\Omega}&=&\int_{\mathbb{R}^3} \bm{\omega}\,d^3x\;,\\
  \bm{I}&=&\frac{1}{2}\int_{\mathbb{R}^3} \bm{x}\times \bm{\omega}\,d^3x\;,\\
  \bm{M}&=&\frac{1}{3}\int_{\mathbb{R}^3} \bm{x}\times\left( \bm{x}\times \bm{\omega}\right)\,d^3x\;.
 \end{eqnarray}
Restricting the integrals to the inner region of self-similar flow, one can easily show, using the axisymmetry and parity conditions of the flow, that all these quantities vanish identically, independently of time and the value of self-similar exponent $\nu$.
 
\begin{figure}
\centerline{  \includegraphics[height=5cm]{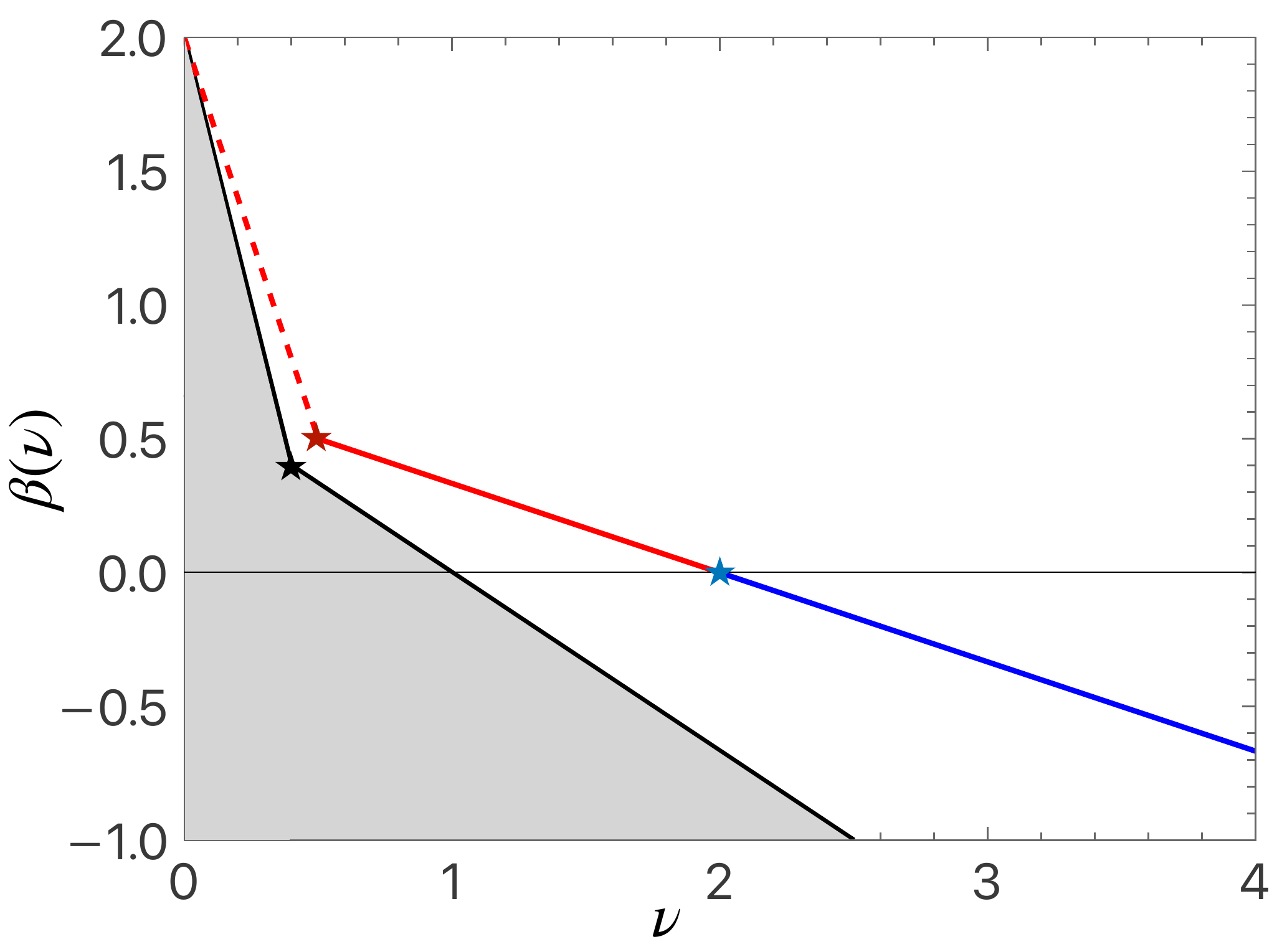}}
\caption{ \label{Fig:phase} A $\beta-\nu$ diagram derived from kinetic energy and helicity conservation laws. The properties of the total kinetic energy require that $\beta(\nu)$ cannot take values in the shaded  region. The total helicity requires that $\beta(\nu)$ takes values given by the red and blue lines. This curve lies in the allowed white region imposed by energy constraints. The dotted line corresponds to the first equality of Eq.~(\ref{eq:selec}) and the solid line to the second equality of equation Eq.~(\ref{eq:selec}). The red colour indicates positive values of $\beta(\nu)$ and blue colour is for $\beta(\nu)<0$. Finally, the three special points $(\nu,\beta)=(1/2,1/2)$, $(2,0)$, $(2/5,2/5)$ are highlighted by a red, blue and black star, respectively.}
\end{figure}

The second group of conservation laws involves scalar physical quantities: the kinetic energy $\mathcal{E}$ (\ref{eq:Energy}) and the helicity $\mathcal{H}$ (\ref{eq:Helicity}) given by
  \begin{eqnarray}
 \mathcal{E}&=&  \mathcal{E}_{\mathrm{\rm in}}(t)+ \mathcal{E}_{\mathrm{out}}(t)=\frac{1}{2}\int_{\mathbb{V}_{\rm in}} \bm{u}^2\,d^3x+ \frac{1}{2}\int_{\mathbb{V}_{\rm out}} \bm{u}^2\,d^3x\, , \label{eq:splittedEnergy} \\
  \mathcal{H}&=& \mathcal{H}_{\mathrm{\rm in}}(t)+ \mathcal{H}_{\mathrm{out}}(t)=\int_{\mathbb{V}_{\rm in}} \bm{u}\cdot\bm{\omega}\,d^3x +  \int_{\mathbb{V}_{\rm out}} \bm{u}\cdot\bm{\omega} \,d^3x\, .  \label{eq:splittedHelicity}
  \end{eqnarray}
  The flow must preserve energy and helicity:  $\mathcal{E}=\mathcal{E}_0$ and $\mathcal{H}=\mathcal{H}_0$ where $\mathcal{E}_0$ and $\mathcal{H}_0$  are the initial kinetic energy and the helicity. Using the results of the {\it swirled} solution, the inner kinetic energy of the flow (\ref{eq:splittedEnergy}) is given by 

\begin{equation}
\frac{t_c^2}{R_0^5}\,{\mathcal E}_{\mathrm{\rm in}}(t) =\frac{L(t)}{R_0}\left(\frac{t_c-t}{t_c}\right)^{4\nu-2} E_1(\nu) +\left(\frac{L(t)}{R_0}\right)^3\left(\frac{t_c-t}{t_c}\right)^{2\nu-2} E_2(\nu) \;,
   \end{equation}
   with
    \begin{eqnarray} 
 E_1(\nu)&=& 2\pi \int_{0}^{1}   \xi^3 U_r ^2( \xi ) d\xi \nonumber\\
    && + \frac{\pi}{2} \left( (a + \nu)  \left((a +\nu) \xi_{\mathrm{in}}^2 -3a+\nu \right)\left(\xi_{\mathrm{in}}^2-1\right) +
    4a^2 \log \xi_{\mathrm{in}}\right) \;,\label{eq:E1}\\
E_2(\nu)&=& \frac{2\pi }{3}   \int_0^{1}   \left(U_z^2(\xi)+S_\phi(\xi)\right) \xi d\xi + \frac{4\pi }{3}  (a + \nu)^2 \left( \xi_{\mathrm{in}}^2-1\right)\;,\label{eq:E2}
    \end{eqnarray}
where $\xi_{\rm in} = R_{\rm in}(t)/ R(t)\simeq \rm Cte >1$, and $a$ is given by Eq.~(\ref{eq:valueOfa}). Notice that $E_1(\nu)>0$ and $E_2(\nu)>0$ ensuring that both terms contribute additively to the energy and cannot cancel each other. The outer flow within $\mathbb{V}_{\rm out}$ carries a kinetic energy ($\mathcal{E}_{\mathrm{out}}(t)\neq0$). Therefore, to ensure a constant total kinetic energy $\mathcal{E}=\mathcal{E}_0$, the energy $\mathcal{E}_{\mathrm{\rm in}}(t)$ must  be finite but not necessarily time-independent. To satisfy this condition, one should impose the following conditions 
\begin{equation}
\frac{L(t)}{R_0}\left(\frac{t_c-t}{t_c}\right)^{4\nu-2} \rightarrow C_1 \quad \& \quad\left(\frac{L(t)}{R_0}\right)^3\left(\frac{t_c-t}{t_c}\right)^{2\nu-2} \rightarrow C_2  \quad \mathrm{as}\quad t\to t_c\;,
   \end{equation}
where $C_1\geq 0$ and $C_2\geq 0$ are constants.
Therefore, if one writes the size of the cylindrical region where the self-similar behaviour holds as:
\begin{equation}
 L(t)\sim R_0\left(\frac{t_c-t}{t_c}\right)^{\beta(\nu)}\quad \mathrm{as}\quad t\to t_c\;,
     \end{equation}
     then the conditions
   \begin{equation}
\beta(\nu)\geq2-4\nu \quad \& \quad \beta(\nu)\geq\frac{2}{3}(1-\nu) \;, \label{eq:betaEnergy}
   \end{equation}
   should be satisfied. Which means that the values of $\beta(\nu)$ in the shaded region of Fig.~\ref{Fig:phase} are not allowed.

For the helicity, the calculations are simple. By using the fact that for $r> R_{\mathrm{in}}(t)$, the flow satisfies ${\bm\omega}=0$, one automatically has
\begin{equation}
\mathcal{H}\equiv \mathcal{H}_{\rm in}(t)=\mathcal{H}_0\;.
   \end{equation}
On the other hand, the helicity (\ref{eq:Helicity}) follows  a similar scaling law as for the energy. One easily shows that
\begin{equation}
 \frac{t_c^2}{R_0^4}\,{\mathcal H}_{0} =\frac{L(t)}{R_0}\left(\frac{t_c-t}{t_c}\right)^{3\nu-2} H_1(\nu) +\left(\frac{L(t)}{R_0}\right)^3\left(\frac{t_c-t}{t_c}\right)^{\nu-2} H_2(\nu) \;,
\label{selecH}
   \end{equation}
with
\begin{eqnarray} 
H_1(\nu)&=&- 4\pi  \int_0^1    \xi^2 U_r \left( \xi \right) U_\phi \left( \xi \right) d\xi >0\;, \\
H_2(\nu)&=& - \frac{8\pi }{3}   \int_0^1  \xi  U_z'(\xi) U_\phi \left( \xi  \right)  d\xi>0\;.
\end{eqnarray}
Notice that the helicity is focused inside the column delimited by $R(t)$, it does not depend explicitly on $\xi_{\rm in}$. Moreover, for the self-similar Ansatz \ref{Secc:Ansatz1}, the integrals $H_1(\nu)$ and $H_2(\nu)$ are finite and strictly positive for all $\nu$.~\footnote{We recall that for $0\leq\xi\leq1$, $U_r(\xi)<0$, $U_\phi (\xi)\geq 0$ and $U_z'(\xi) < 0$ (See Fig. \ref{Fig:ODEsSolution}).} This ensures that both terms contribute additively to the total helicity and do not cancel each other \footnote{{\it A priori} the helicity may also be negative, by simply changing the parity of the flow.}. Because the helicity must be finite for all times, Eq. (\ref{selecH}) imposes the dynamics of stretching $L(t)$. Therefore, we conclude that a satisfactory solution  requires for $t\rightarrow t_c$
\begin{equation}
\frac{L(t)}{R_0}\left(\frac{t_c-t}{t_c}\right)^{3\nu-2} \sim 1 \quad \& \quad \left(\frac{L(t)}{R_0}\right)^3\left(\frac{t_c-t}{t_c}\right)^{\nu-2} \rightarrow C_3 \;,
   \end{equation}
or
\begin{equation}
\frac{L(t)}{R_0}\left(\frac{t_c-t}{t_c}\right)^{3\nu-2} \rightarrow C_4 \quad \& \quad \left(\frac{L(t)}{R_0}\right)^3\left(\frac{t_c-t}{t_c}\right)^{\nu-2} \sim 1 \;,
   \end{equation}
 where $C_3\geq0$ and $C_4\geq0$ are constants. This yields to the following condition
     \begin{equation}
     \beta(\nu)=\left\{
     \begin{array}{ll}
\displaystyle{2-3\nu} & \displaystyle{\nu\leq{1}/{2}}\\
\displaystyle{ \frac{2-\nu}{3}} &\displaystyle{ \nu\geq{1}/{2}}
\end{array}
\right.\;.
\label{eq:selec}
   \end{equation}
As shown in Fig.~\ref{Fig:phase}, the behaviour of $\beta(\nu)$ as given by (\ref{eq:selec}) lies in the allowed (white) region derived from the kinetic energy considerations. An important consideration is that the helicity bounds for $\beta(\nu)$ does not cross the energy bounds for the exponent. Moreover, notice that $\beta(\nu<2)>0$ and $\beta(\nu>2)<0$. Therefore, for $\nu<2$ it is expected that the blowup induces a point-like singularity, while for $\nu\geq 2$ we recover essentially Pomeau's scenario for which the finite-time singularity is along a line. 

Up to this point, we have identified a continuous family of possible solutions characterised by $\beta(\nu)$. However, a selection mechanism for $\nu$ is still required, which calls for additional physical constraints. We can proceed further by introducing physical arguments to single out particular values, and show that $\beta(1/2)=1/2$ and $\beta(2)=0$ are two plausible candidates. For a point-like singularity, one expects $L(t)\sim R(t)$, since the orientation of cylindrical coordinates relative to any Cartesian reference frame should not affect the dynamics. This assumption leads to $\nu=1/2$, which corresponds to the original exponent proposed by \cite{leray1934} for the Navier-Stokes equations. On the other hand, when $\nu>2$, one expects that $L(t)$ should not exceed the characteristic length scale $\sim R_0$ set by the initial flow. Therefore, one may assume $L(t)\sim R_0$, which formally leads to the marginal value $\beta\to 0$ and thus to $\nu \to 2$.

 \subsection{The {\it swirless} zero-helicity case}
 \label{sec:swirless}
 
In this Section, we discuss the physical implications of the solution without swirl (\ref{eq:UrUzZeroSwirl}). One can follow the same approach as in the {\it swirled} case, with the main difference that the total helicity vanishes identically,  while, the kinetic energy remains a nonzero conserved quantity. The main difference is that $E_1(\nu)$ and $E_2(\nu)$ given by Eq.~(\ref{eq:E1}) and Eq.~(\ref{eq:E2}) respectively should be replaced by
    \begin{equation} 
 E_1(\nu)= 2\pi \int_{0}^{\xi_{\mathrm{in}}}   \xi^3 U_r ^2( \xi ) d\xi \quad\&\quad
E_2(\nu)= \frac{2\pi }{3}   \int_0^{\xi_{\mathrm{in}}}   \xi U_z^2(\xi)  d\xi\;,
    \end{equation}
    with $U_r(\xi)$ and $U_z(\xi)$ given by Eq.~(\ref{eq:UrUzZeroSwirl}).
In the {\it swirled} case, the kinetic energy constraints impose that the allowed solutions for $\beta(\nu)$ lie, precisely, in the white region of Fig.~\ref{Fig:phase} including the solid black curve.

\begin{figure} 
\centerline{\includegraphics[width=12cm]{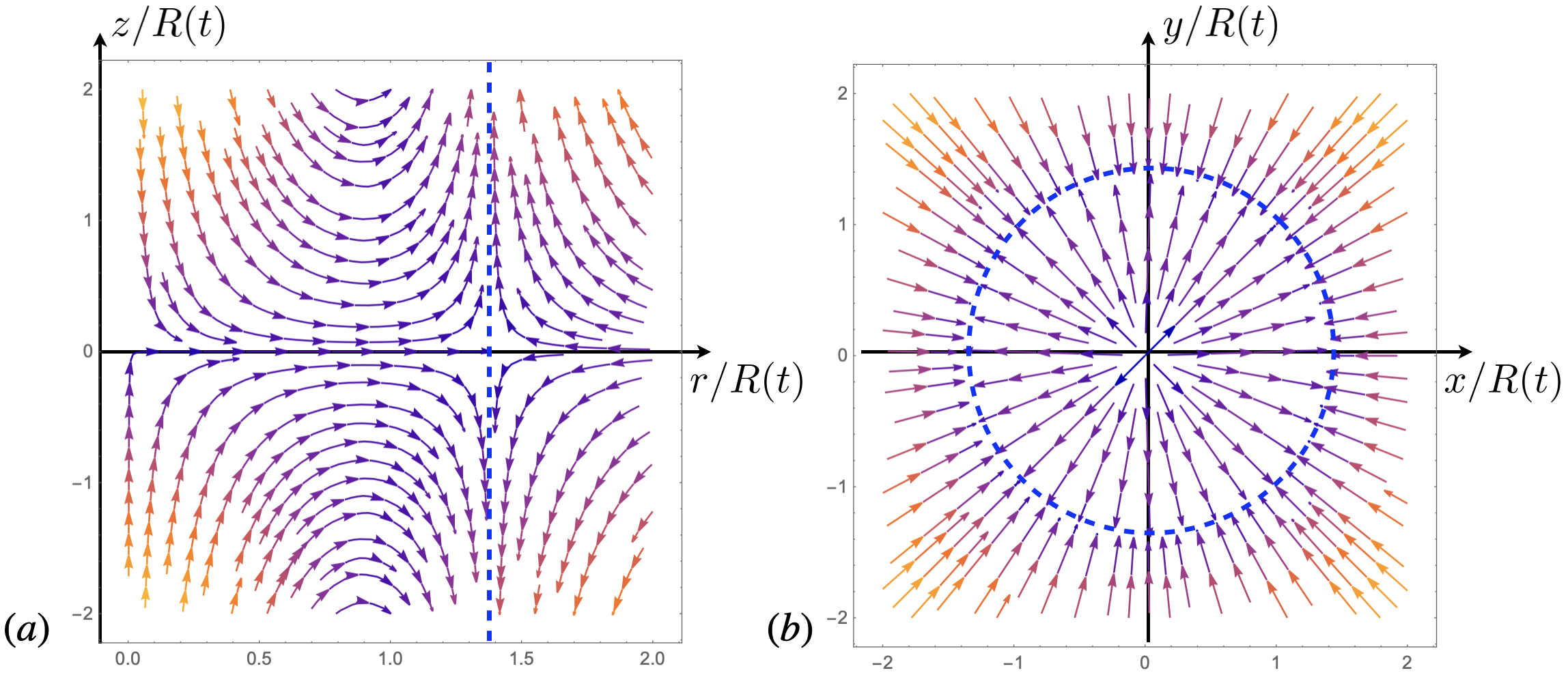}}
\caption{ \label{Fig:StreamSwiless}  The streamlines of the {\it swirless} case for $\nu=2$ corresponding to the flow of the scaled axisymmetric velocity field $\frac{(t_c-t)}{R(t)}\,\bm{u}(\bm{x},t)$ as function of the dimensionless coordinates $(r/R(t),z/R(t))$. Panel (a) shows the streamlines in the $(r,z)$-plane, and, panel (b) shows the streamlines in the $(x,y)$-plane at $z=0$. The dotted blue line and the circle correspond to the surface $U_r(\xi_0)=0$ where the streamlines change direction from outward ($\xi<\xi_0$) to inward ($\xi>\xi_0$). }
\end{figure}

Interestingly, the inspection of the streamlines of the {\it swirless} solution projected in $(r,z)$-plane show no qualitative differences from the {\it swirled} case: both $U_r(\xi)$ and $U_z(\xi)$ vanishes at some $\xi$ creating a stagnation point on the plane $z=0$ (See Fig. \ref{Fig:StreamSwiless}-(a)). However, in the {\it swirled} case the flow is outward the $z$-axis, while, it is inward in the {\it swirless} case implying that the {\it swirless} case does not develop a focusing mechanism. On the contrary, the $(x,y)$-projection shows more relevant difference among the two cases. In particular, it is visible that the {\it swirled} case involves a rotation of the flow (see Fig.~\ref{Fig:Stream2D}(b)), whereas in the {\it swirless} case, the flow corresponds to a radial flow with a stagnation point at the origin (see Fig. \ref{Fig:StreamSwiless}(b)).

The obtained solutions show that existence of a blow-up could be associated to the existence of swirl and the {\it helicity} emerges as the relevant parameter and not the kinetic energy. We conjecture that if the initial helicity vanishes, then, the blow-up of solutions of Euler equations at finite time are suppressed. This is supported by a simple dimensional analysis argument. After Eq.~(\ref{eq:scales}), the critical time $t_c$ and lengthscale $R_0$ scale as :
\begin{equation}
t_c \sim    \frac{\mathcal E_0^{2}}{\mathcal H_0^{5/2}} \quad\&\quad R_0\sim \frac{\mathcal E_0}{\mathcal H_0} \;.
\label{eq:scales1}
\end{equation}
Therefore, as $\mathcal H_0\to 0$, the time scale $t_c$, as well as, the length $R_0$ diverge, calling for a redefinition of $R(t)$ as given by Eq.~(\ref{eq:SelfSimilarRadiius}). Using Eq~(\ref{eq:scales1}), then the self-similar length-scale (\ref{eq:SelfSimilarRadiius}) yields~:
\begin{equation}
R(t)=R_0t_c^{-\nu}(t_c-t)^\nu\sim \frac{\mathcal E_0^{(1-2\nu)}}{\mathcal H_0^{(1-5\nu/2)}}(t_c-t)^\nu\;.
\label{eq:SelfSimilarRadiius2}
\end{equation}
From (\ref{eq:SelfSimilarRadiius2}) it is clear that to keep the self-similar behaviour of the {\it swirless} solution for $\mathcal H_0\rightarrow 0$, it is necessary to impose $\nu=2/5$ to reach a finite scaling law for $R(t) \sim { \mathcal E_0}^{1/5} (t_c-t)^{2/5}$, which corresponds to the Taylor-Sedov-von Neuman scaling \citep{Taylor1950a,Sedov1946,vonNeuman}. The exponent $\nu=2/5$ was investigated by \cite{yves2018} in connection with Euler equations. In addition, using the energy conditions (\ref{eq:betaEnergy}) or the information in the Fig.~\ref{Fig:phase}, we can show that for $\nu=2/5$, then $\beta=2/5$.

In conclusion, our findings indicate that for a {\it swirless} flow, if a point-like singularity for solutions of the Euler equations with a self-similar exponent $\nu=\beta=2/5$ do exist, then this singularity occurs in infinite time.

\section{Discussion}

We have identified a self-similar Ansatz for the Euler equations that admits semi-analytical solutions and leads, in the presence of swirl, to a finite-time singularity under minimal assumptions.
Our results are sufficiently general as we conjecture they are valid for some set of smooth initial condition of the inviscid flow in ${\mathbb R}^3$ provided it satisfies regularity and convergence conditions at infinity. The obtained solutions are only characterised by the conserved quantities, namely  the total kinetic energy and helicity which are conserved through time during the evolution of the flow towards the singularity.  The initial energy  $\mathcal{E}_0$ is, by definition, strictly positive while the helicity is an arbitrary  real number. These two global quantities of the flow allow for the definition of a lengthscale $R_0$ and a timescale $t_c$ which may be identified as the critical time and are used to build the self-similar Ansatz. Since Euler equations do not involve any physical scale, $R_0$ and $t_c$  are necessarily given by the characteristics of the initial flow ${\bm u}_0(\bm{x})$ at $t=0$.

In this paper we studied two distinct type of solutions~: a structureless {\it swirless} solution with $\mathcal{H}_0=0$, and, a {\it swirled} solution with $\mathcal{H}_0\neq0$ that presents a self-focusing of helicity along a tubular region around the $z$-axis. The {\it swirless} solution does not exhibit any blow-up in finite time. On the contrary, since the {\it swirled} solution preserves the total helicity within $0<t<t_c$, it allows for finding the dynamics of the relevant length scales, $R(t)$ and $L(t)$, and develops a blow-up at finite time $t=t_c$. Moreover, two distinct blow-up regimes arise: a point-like and a line-like finite-time singularities.

The most peculiar property of the obtained {\it swirled} solution is that it may be interpreted a {\it two-fluid flow} featured by the coexistence of a {\it swirled} and a {\it swirless} flow separated by a sharp interface of discontinuity with an infinite gradient of the swirl velocity. Indeed, the swirl velocity and the vorticity field are compact functions that vanishes at a finite distance, $R(t)$, in finite time. This process appears to be very appealing because it defines a geometrical interface that shrinks into a line (or a point as discussed) in finite time. More importantly, using additional physical arguments we are able to select the exponent $\nu$ for the {\it swirled} solutions. For a point-like singularity (a singularity of co-dimension $0$) the self-similar exponent becomes $\nu=1/2$ (and $\beta=1/2$), which corresponds to the  Leray exponent for Navier-Stokes equations. On the other hand, for a line-like singularity (a singularity of co-dimension $1$), the self-similar exponent is $\nu=2$ (and $\beta=0$). Incidentally, the semi-theoretical calculation based on a multipolar expansion of Euler equations also predicts $\nu=2$ \citep{PRF2024Diego}.

Concerning the {\it swirless} flow, by using dimensional scaling arguments we found that the self-similar exponents should be: $\nu=\beta=2/5$ near the singularity, but this singularity occurs at infinite time from the initial condition. Moreover, there is no a compact structure as in the {\it swirled} case. In the {\it swirless} case,  we are in presence of a smooth solution as time goes by.

Finally, we end this paper with the following closing remarks:
\begin{itemize}
\item The solutions obtained in Section~\ref{Secc:quatre} admits a large (probably infinite) number of solutions depending on the selection of the value of $a_0$, which we limited only to two types of solutions. In a future work we will extend this study to exploring others solutions. 
\item  The current approach paves the road to the problem of the existence of a point-like singularity in the Navier-Stokes equations. The Leray singularity mandates a self-similar exponent $\nu=1/2$, therefore the obtained point-like finite-time singular solution for perfect fluids with $\nu=\beta=1/2$ may be relevant for the viscous case. We expect that the diffusion term of Navier-Stokes equations may act as a regularising mechanism for the sharp interface separating the {\it swirled} and {\it swirless} phases rendering the component $u_\phi$ of the velocity field more regular than being only of class ${\mathcal C}^0$. Indeed, it looks straightforward to generalise Sections~\ref{Secc:Formulation} and \ref{Secc:Ansatz} to Navier-Stokes equations. Moreover, it is possible to show that there exists a {\it swirless} solution similar to the one obtained in  Section~\ref{Secc:SwirlessSolution}, and, we believe we can build an equivalent numerical scheme to look for possible {\it swirled} solutions. However, as our analysis of the solutions to Euler equations prior to blow-up is based on conservation laws, this approach should be modified for the viscous case. Conservation laws for Navier-Stokes equations include a dissipation rate, and, therefore the main challenge is to build a physical approach to look for possible finite-time singularities where both injected energy and helicity and their viscous dissipative counterparts enter in play.
\end{itemize}


\bibliographystyle{jfm}

\end{document}